\documentclass[%
 reprint,
 superscriptaddress,
nofootinbib,
 amsmath,amssymb,
 aps,
]{revtex4-2}

\usepackage{amsmath}
\usepackage{amssymb}
\usepackage{graphicx}
\graphicspath{{../Figures/}{../}}
\usepackage{dcolumn}
\usepackage{bm}
\usepackage{mathptmx} 
\usepackage{hyperref}
\usepackage{xcolor}
\usepackage{bbold}
\definecolor{darkblue}{rgb}{0, 0.4, 0.6}
\hypersetup{
    colorlinks=true,
    linkcolor=darkblue,
    filecolor=darkblue,      
    urlcolor=darkblue,
    citecolor = darkblue,
    pdftitle={Witzany et al.: Actions of spinning compact binaries},
    pdfpagemode=FullScreen,
    }
\usepackage{orcidlink}
\usepackage{fancyhdr}
\usepackage{extramarks}
\makeatletter

\renewcommand{\d}{\mathrm{d}}

\newcommand{\sn}{\,\mathrm{sn}\,}
\newcommand{\cn}{\,\mathrm{cn}\,}
\newcommand{\dn}{\,\mathrm{dn}\,}

\begin{document}

\preprint{APS/123-QED}

\title{Actions of spinning compact binaries: Spinning particle in Kerr matched to dynamics at 1.5 post-Newtonian order}

\author{Vojtěch Witzany \orcidlink{0000-0002-9209-5355}}%
 \email{vojtech.witzany@matfyz.cuni.cz}
\affiliation{%
 Institute of Theoretical Physics,
Charles University, V Holešovičkách 2,
180 00, Praha 8, Czech Republic 
}%
\author{Viktor Skoupý \orcidlink{0000-0001-7475-5324}}%
\affiliation{%
 Institute of Theoretical Physics,
Charles University, V Holešovičkách 2,
180 00, Praha 8, Czech Republic 
}%
\author{Leo C. Stein \orcidlink{0000-0001-7559-9597}}
\affiliation{
 Department of Physics and Astronomy, University of Mississippi, University, MS 38677, USA
}%
\author{Sashwat Tanay \orcidlink{0000-0002-2964-7102}}
\affiliation{
 Laboratoire Univers et Théories, Observatoire de Paris, Université PSL, Université Paris Cité,
CNRS, F-92190 Meudon, France
}%

\date{\today}

\begin{abstract}
The motion of compact binaries is influenced by the spin of their components starting at the 1.5 post-Newtonian (PN) order. On the other hand, in the large mass ratio limit, the spin of the lighter object appears in the equations of motion at first order in the mass ratio, coinciding with the leading gravitational self-force. Frame and gauge choices make it challenging to compare between the two limits, especially for generic spin configurations. We derive novel closed formulas for the gauge-invariant actions and frequencies for the motion of spinning test particles near Kerr black holes. We use this to express the Hamiltonian perturbatively in terms of action variables up to 3PN and compare it with the 1.5 PN action-angle Hamiltonian at finite mass ratios. This allows us to match the actions across both systems, providing a new gauge-invariant dictionary for interpolation between the two limits. 
\end{abstract}

\maketitle
\onecolumngrid
\begin{center}
\vskip -2 em
\par\noindent\rule{0.95\textwidth}{0.5pt}
\end{center}
\vskip -2 em
\tableofcontents
\begin{center}\par\noindent\rule{0.95\textwidth}{0.5pt}\end{center}
\vskip 1 em
\twocolumngrid

\section{Introduction} \label{sec:intro}

The relativistic two-body problem -- and the relativistic problem of motion in general -- are notoriously difficult, which was realized soon after the publication of the Einstein equations by the likes of Einstein, Lorentz, Droste, Infeld and Hoffmann \cite{lorentz1937motion,Einstein:1938yz,1987thyg.book..128D}. Due to the non-linear nature of the Einstein equations, the field problem has to be tackled simultaneously with the motion of the sources. With physical boundary conditions, eternal bound motions do not exist; the binary systems emit gravitational waves, leading to orbital decay and eventual merger. Consequently, one often resorts to various analytical weak-coupling expansions \cite{Blanchet:2013haa,poisson2014gravity,Barack:2018yvs,Pound:2021qin} or complex numerical-relativity setups to solve for the evolution \cite{baumgarte2010numerical,LISAConsortiumWaveformWorkingGroup:2023arg}. Furthermore, even for the simplest binary objects, such as black holes, one must consider not only the motion of their centers of mass but also their spin degrees of freedom \cite{Damour:2001tu,schafer2009gravitomagnetism}. 

The historically first weak-coupling approach to the relativistic two-body problem is the post-Newtonian (PN) expansion, which deals with systems in the limit of low-velocities $v$ and large-separations $r$ \cite{Damour:1988mr,Blanchet:2013haa,poisson2014gravity}. It is optimal for bound systems where the virial theorem indicates that the dimensionless velocity squared scales inversely with separation $v^2/c^2 \sim GM/(r c^2)$, with $c$ being the speed of light, $G$ Newton's constant, and $M$ a characteristic mass. For compact binaries, it also allows to iterate from the well-understood Newtonian two-body problem. 

Specifically, the Newtonian problem of two compact objects represented as point masses has 6 degrees of freedom corresponding to their spatial positions. In PN theory, each binary member gains an additional spin degree of freedom, yielding a total of 8 degrees of freedom \cite{Damour:2001tu,schafer2009gravitomagnetism}. This indicates that the dynamical system is not necessarily integrable based solely on the number of integrals of motion derived from the system's rotational and translational symmetries. In general, these symmetries guarantee 6 mutually independent conservation laws, so another 2 have to be supplied for full integrability \cite{Damour:2001tu,Tanay:2020gfb}. 

Another weak-coupling approach suitable for bound systems is the post-geodesic or self-force expansion \cite{Barack:2018yvs,Pound:2021qin}. This method starts with the motion of a free test particle, representing the lighter secondary member of the binary, in a given strong gravitational field such as that of an isolated Kerr black hole, representing the field of the heavier primary. The motion is then iterated by treating the ``particle" as a small mass that sources a metric perturbation, causing deviation from the original geodesic; this is the gravitational self-force. Even though the trajectory and the metric perturbation satisfy non-trivial field and evolution equations on a fully relativistic background, the motion and field equations decouple into a more manageable iterative scheme within the self-force expansion. This scheme is particularly effective for binaries with extreme mass ratios \cite{vandeMeent:2020xgc}. However, due to label-switching symmetries between the two bodies \cite{Damour:2016gwp}, its results can also describe binaries of comparable masses unexpectedly well \cite{LeTiec:2011bk,Pound:2019lzj,vandeMeent:2020xgc,Warburton:2021kwk,Wardell:2021fyy}.

How do the two aforementioned approximations relate when dealing with spins in the binary? In the PN scheme, the spins of the binary members are represented by two Cartesian-like 3-vectors $\vec{S}_{1,2}$ attached to the center of mass of each body \cite{schafer2009gravitomagnetism,Steinhoff:2011sya}. Their dynamics are trivial at leading Newtonian order (the spin vectors are constant) and start to couple to the system only at the 1.5 PN order ($ \mathcal{O}(v^3/c^3)$ relative to Newtonian order). The dynamics are cast in a global inertial frame and both spins evolve. In contrast, in the post-geodesic expansion, each spin is treated very differently \cite{Pound:2021qin,Witzany:2018ahb}. The primary is so heavy that the inertial frame becomes fixed to its spin at zeroth order, and any dynamics of the primary spin vector are erased from the phase space. The spin of the secondary is represented by a covariant 4-vector $s^\mu$, fulfilling non-trivial parallel transport equations at zeroth order and back-reacting on the lighter object's trajectory in the first-order mass ratio iteration. At first glance, relating the two formalisms is not straightforward.

Capturing the invariant consequences of spin is challenging. The center of mass of spinning bodies is not defined uniquely. Different parts of rotating objects exhibit varying gravitational and/or inertial masses in different frames \cite{1965PhRv..137..188F}. Typically, this frame is determined by a supplementary spin condition \cite{Costa:2014nta}. In numerical relativity, the computation of spin and center of mass is more direct and relies on the time slicing and a chosen quasi-local prescription for computation \cite{Lovelace:2008tw,Owen:2017yaj}. Similar to gauge choices in field theory, these various options yield physically equivalent descriptions of spinning binaries, differing only in superficial variables \cite{Costa:2014nta,Steinhoff:2015ksa}. Moreover, the variables describing the motion can carry ``pure gauge'' degrees of freedom, causing the system to exhibit variability at new ``Zitterbewegung'' frequencies unrelated to the physical frequencies of the system (see e.g. \cite{Costa:2011zn}). 

We need to resolve these issues soon. With the increasing sensitivity and reach of current and near-future gravitational wave detectors \cite{abbott2020prospects,Reitze:2019iox,Punturo:2010zz,TianQin:2015yph,LISA:2017pwj}, we will observe new types of binary inspirals with different mass ratios, spin configurations, and eccentricities. Maximizing the scientific yield of such observations necessitates significant improvements in the modeling of binary inspirals and the emergent waveforms \cite{LISAConsortiumWaveformWorkingGroup:2023arg,Jan:2023raq}. Further synergies among the various approaches listed above are also necessary. Establishing reliable links between PN, numerical relativity, and self-force formalisms will require focusing on carefully chosen observables, or gauge invariants, which is the focus of this work. 

How can we link the large mass ratio and PN pictures with general spin dynamics? At leading order, the dynamics are fully conservative \textit{and integrable} both in the 1.5PN limit \cite{Damour:2001tu,Tanay:2020gfb,Tanay:2021bff} and the spinning test particle limit \cite{Witzany:2018ahb,Witzany:2019nml}. This allows a description via the geometric formalism of Hamiltonian mechanics of integrable systems \cite{Arnold:1989who,Arnold:2006}. Specifically, we can compute the so-called \textit{actions} of the system with nearly ``topological'', gauge-invariant properties. Among other features, determining their functional form allows us to compute the fundamental frequencies of motion \cite{Schmidt:2002qk}, an essential observable in gravitational wave science. 

Importantly, one set of actions can transform into another \textit{ only} via a discrete lattice transform that corresponds to switching between inequivalent homotopy classes of curves in phase space \cite{Arnold:1989who,Arnold:2006,fasano2006analytical}. When computing actions in continuous parameter limits, they must reduce to each other up to discrete transforms. This is how we establish the correspondence between the post-geodesic and PN limits in this paper. 

Indeed, the direct matching of action variables between the Hamiltonian for nonspinning test particle motion in an effective metric and a PN Hamiltonian for nonspinning binaries was the one of the two main ingredients of the celebrated Effective one-body formalism \cite{Buonanno:1998gg}. The original Effective one-body model by \citet{Buonanno:1998gg} matched the actions of the PN dynamics with those of a nonspinning test particle in an effective deformed Schwarzschild metric. The second step was identifying optimal candidates for matching the \textit{parameters} of the two systems. The authors required the mass of the test particle $m$ and Schwarzschild-like mass $M$ to match the reduced mass $m_1 m_2/(m_1+m_2)$ and total mass $m_1 + m_2$ of the binary. This led to a mapping between the energies of both systems and a prescription for the effective metric in the test particle Hamiltonian. Consequently, this approach achieved a highly efficient resummation of the PN dynamics, smoothly connecting the plunge and merger phases to the inspirals, and thus provided one of the first qualitatively accurate models of the full coalescence process \cite{Buonanno:1998gg,Buonanno:2000ef}. 

The effective one-body formalism was later modified to include spin effects through various different approaches. Damour incorporated spins of the binary components by replacing the Kerr spin vector $M a \hat{z}$ with a non-trivial linear combination of $\vec{S}_1$ and $\vec{S}_2$ in the test particle Hamiltonian in a deformed Kerr metric \cite{Damour:2001tu}. Barausse \& Buonanno offered an alternative approach, substituting the Kerr spin vector at leading order with $\vec{S}_1 + \vec{S}_2$, and endowing the test particle with a leading-order spin $\sim m_2/m_1 \vec{S}_1 + m_1/m_2 \vec{S}_2$ \cite{Barausse:2009xi}. Both approaches matched spin-like expressions with \textit{different} Poisson brackets in each of the two systems. As such, the various spin-like vectors have different equations of motion on each side of the correspondence. In other words, the correspondence is well suited for the case of fully aligned spin vectors without precession, where the spin degrees of freedom are inactive. However, it cannot be expected to work well once the spins become dynamical. One of our goals in this paper is to also provide a basis for an improved Effective one-body formalism for spinning binaries in general dynamical states.

\section{Summary of results} \label{sec:summary}

\begin{itemize}
    \item We derive novel closed-form expression for the action variables of spinning test particles in Kerr space-time to linear order in the particle spin. The results are expressed using Legendre elliptic integrals and apply to general bound motion (arbitrary eccentricity and spin-to-orbit inclinations). Even the closed expressions for the geodesic limit of the actions (zero spin case) were not available in the literature before this work.
    \item Using the actions, we employ the method of \citet{Schmidt:2002qk} to also obtain closed-form expressions of the fundamental frequencies of motion of spinning test particles in Kerr space-time for the first time. They agree with previous results obtained using numerical and semi-analytical methods (see Fig. \ref{fig:comparisons_DH}).
    \item We also derive the expression for the Hamiltonian of the spinning particle in terms of actions valid up to 3PN order and generic bound orbits. We match the results to the 1.5PN action-angle formalism for spinning finite-mass ratio binaries of \citet{Tanay:2020gfb,Tanay:2021bff} in the spinning test particle limit. In passing, we provide the first closed-form expression for the 1.5PN Hamiltonian in terms of actions.
    \item The match reveals a novel dictionary between the degrees of freedom of spinning binaries in the test particle and weak-field limits. It also opens issues to be addressed by future work. 
\end{itemize}

\section{Organization and notation} \label{sec:organization}
In Section \ref{sec:sptp} we summarize the solution of the Hamilton-Jacobi equation for spinning particles in Kerr space-time. We then use it to compute the corresponding actions and frequencies. In Section \ref{sec:match} we derive the matching of the 1.5PN actions of spinning binaries at finite mass ratio and those of the spinning particle. We do so by expanding the expressions for the spinning-particle actions and then inverting for energy (the Hamiltonian). Conversely, we also reduce the 1.5PN Hamiltonian at finite mass ratio to the spinning test particle limit and find the transformation between the two sets of actions. Section \ref{sec:discussion} discusses the implications of this work and possible future developments.

We use the $G=c=1$ units and $(-+++)$ metric signature. Greek indices are space-time indices, large Latin indices are tetrad indices, and the semi-colon denotes covariant differentiation.

\section{Actions and frequencies for spinning particle in Kerr} \label{sec:sptp}

The motion of an extended test body in curved space-time is described in a multipolar expansion by the so-called Mathisson-Papapetrou-Dixon (MPD) equations \cite{Mathisson:1937zz,Papapetrou:1951pa,Dixon:1970I,Dixon:1970II,DixonIII}. Truncating these at pole-dipole order yields the spinning test particle approximation. This truncation is natural for compact objects moving in an ambient space-time\footnote{The original MPD equations were derived for material bodies, but they can be proven to hold at least to pole-dipole order also for the evolution of spinning black holes in ambient curved space-time \cite{1975PhRvD..11.1387D,Thorne:1984mz}.} since the multipolar expansion is controlled by the ratio between the object's physical radius to the ambient curvature radius \cite{Steinhoff:2012rw}. In compact, large-mass-ratio binaries, the dipole term becomes a correction in the equations of motion of the order of the small mass ratio. 

In this section, we compute the invariant actions and fundamental frequencies for spinning particles in Kerr space-time to linear order in spin. We introduce the Hamiltonian formalism for spinning particle motion and summarize the Hamilton-Jacobi equation solution from Ref. \cite{Witzany:2019nml}, emphasizing aspects relevant to the action computation. The actions are derived via an analytical-extension argument and converted to Legendre form. Finally, fundamental frequencies are obtained from derivatives of the actions using formulas similar to those of \citet{Schmidt:2002qk} for Kerr geodesics.

\subsection{The Hamiltonian formalism for spinning particles}

Within the Hamiltonian formalism of \citet{Witzany:2018ahb}, one can cover the phase space of a spinning test particle with canonical coordinates and obtain the Hamiltonian governing the motion truncated to linear order in spin as
\begin{align}
\begin{split}
H_{\rm TD} = 
    & \frac{1}{2} g^{\mu\nu}\! \left(\!\pi_\mu - \frac{1}{2}  {e}_{A;\mu}^\kappa  {e}_{B \kappa} {s}^{AB}\!\right)\! \left(\!\pi_\nu - \frac{1}{2}  {e}_{C;\nu}^\lambda  {e}_{D \lambda} {s}^{CD} \!\right) 
    \\ 
    & + \mathcal{O}(s^2) \,, 
\end{split}
\label{eq:HTD}
\end{align} 
where $\pi_\mu$ are canonical momenta canonically conjugate to the position of the particle $x^\mu$, and $s^{AB} = -s^{BA}$ are components of the specific spin tensor $s^{\mu\nu}$ with respect to some fixed tetrad field $e^A_{\mu}$. The expression ${e}_{A;\mu}^\kappa  {e}_{B \kappa}$ are the spin-connection (or Ricci rotation) coefficients compatible with the Levi-Civita connection of the background. The antisymmetric tensor $s^{AB}$ Poisson-commutes with all other phase-space coordinates and has the Poisson bracket of the generators of the Lorentz group with itself:
\begin{align}
    \{s^{AB},s^{CD}\} = s^{AC} \eta^{BD} + s^{BD} \eta^{AC} - s^{AD} \eta^{BC} -s^{BC} \eta^{AD} \,.
\end{align}
The spin tensor has two invariants (Casimirs), which commute with every component $s^{AB}$: the spin magnitude $s \equiv \sqrt{s^{AB}s_{AB}/2}$ and $s^* \equiv s^{AB} s^{CD} \epsilon_{ABCD}$. The Casimir $s^* = 0$ for physical spin tensors.\footnote{More precisely: The spin tensor can be decomposed into a mass dipole and a proper spin in any frame. If the spin tensor is defined in any one local frame so that $x^\mu(\tau)$ is the center of mass with respect to that frame, then the mass dipole has to vanish in that frame and the spin tensor has to be degenerate. Since $s^*$ is an invariant proportional to the square root of the determinant of $s^{AB}$, it is then equal to zero when calculated in any frame.} The spin magnitude $s$ or $s/r_{\rm c}$, where $r_{\rm c}$ is the curvature radius of the background, can be viewed as the parameter controlling the deviation from geodesic motion \cite{Steinhoff:2012rw,Witzany:2019nml}. The covering of the space of physical spin tensors $s^{AB}$ by canonical coordinates was given by \citet{Witzany:2018ahb}. An important point for later is that the covering strongly depends on the tetrad $e_A^\mu$ defining the tetrad components in the first place.

The subscript ``TD'' in the Hamiltonian \eqref{eq:HTD} stands for the fact that the evolution corresponds to the so-called Tulczyjew-Dixon or ``covariant'' centroid-fixing condition $s^{\mu\nu} (\pi_\mu - {e}_{A;\mu}^\kappa  {e}_{B \kappa} {s}^{AB}/2) = 0$ \citep{tulczyjew1959motion,Dixon:1970I,Witzany:2018ahb}. This condition is satisfied ``coincidentally'' by the evolution generated by the Hamiltonian \eqref{eq:HTD} when initial data that satisfy the condition are supplied. 

The Hamiltonian \eqref{eq:HTD} generates motion parametrized by proper time for the normalization $H_{\rm TD} = -1/2$. Whenever $s^{AB} = 0$ the motion reduces to a geodesic in the background space-time with $\pi_\mu = u_\mu$ the covariant four-velocity. As such, it is an ideal starting point for a perturbative computation in powers of $s^{AB}$. 

Even though in principle one is tempted to expand the Hamiltonian and discard an $\sim s^2$ cross-term involving the square of the Ricci rotation coefficients, we will use a specific tetrad that has $\sim 1/\sqrt{s}$ singularities in the coefficients near the turning point of the motion below, so we have to keep this term (see Section \ref{subsec:hamjac} for more). Another way to view this ``non-truncation scheme'' is to require strict invariance of the motion with respect to the arbitrary choice of the tetrad $e^A_\mu$, which only serves to generate the particular choice of canonical coordinates. This covariance condition leads to the requirement that the expressions depend on the canonical momentum $\pi_\mu$ only through the \textit{covariant} momentum $p_\mu \equiv \pi_\mu - {e}_{A;\mu}^\kappa  {e}_{B \kappa} {s}^{AB}/2$ in any expression of interest (see Ref. \cite{Witzany:2018ahb} for more).   

\subsection{Marck tetrad congruence} \label{subsec:marck}
When the phase-space of the spinning particle is fully covered by canonical coordinates, it is easy to use the Hamiltonian \eqref{eq:HTD} to formulate the Hamilton-Jacobi equation for the spinning particle in any space-time. This was done by one of us in the Kerr metric in Ref. \cite{Witzany:2019nml}. At geodesic order, the appropriate coordinates for separating the Hamilton-Jacobi equation are the Boyer-Lindquist coordinates and their canonically conjugate momenta, as shown by \citet{Carter:1968rr}. The spinning particle possesses additional degrees of freedom with new coordinate sets on $s^{AB}$, where the covering is generated by a tetrad field. Ref. \cite{Witzany:2019nml} used the Marck tetrad construction \cite{marck1983solution} to generate coordinates on $s^{AB}$. The reason for this Ansatz was that at leading order the spin tensor is parallel-transported, and the Marck tetrad separates the parallel-transport equations along geodesics.

Specifically, in Ref. \cite{Witzany:2019nml} a background geodesic \textit{congruence} was chosen so that the corresponding geodesic motion was close to the motion of the spinning test particle in question, and then the corresponding Marck tetrad was generated from it. The resulting tetrad reads\footnote{The ordering of the legs of the tetrad here corresponds to the generation of the tetrad by iterating powers of the Killing-Yano tensor as described in Ref. \cite{Witzany:2019nml}. It disagrees with the ordering and sign convention of \citet{marck1983solution} and \citet{vandeMeent:2019cam} while keeping the same orientation of the basis.}
\begin{subequations}
\label{eq:mtet}
\begin{align}
    & e_\mu^{0} =
        \begin{pmatrix}
        -E_{\rm c} \\
        L_{\rm c} \\
        u_{r \rm c} \\
        u_{z \rm c} 
        \end{pmatrix},
    \\
    & e_\mu^{1} = \frac{1}{ \sqrt{\Lambda K_{\rm c}}}
        \begin{pmatrix}
        \frac{1}{\Sigma}\left(
            a^2 z (1 - z^2) u_{z \rm c} - \Lambda r \Delta u_{r \rm c}
        \right) 
        \\
        \frac{a (1 - z^2)}{\Sigma}\left(
            \Lambda r \Delta u_{r \rm c} - z (r^2 + a^2) u_{z \rm c}    
        \right) 
        \\
        \frac{\Lambda r}{\Delta} \left((r^2 + a^2) E_{\rm c} - a L_{\rm c} \right) \\
        - az \left( a E_{\rm c} - \frac{L_{\rm c}}{1 - z^2}\right) 
        \end{pmatrix},
    \\
    & e_\mu^{2} = 
        \frac{1}{\sqrt{\Lambda}}
        \begin{pmatrix}
        \frac{ (r^2 + a ^2) E_{\rm c} - a L_{\rm c} }{K_{\rm c} + r^2} -E_{\rm c}
        \\
        \Lambda L_{\rm c} + \frac{(r^2 + a^2)(L_{\rm c} -a (1 - z^2) E_{\rm c})}{K_{\rm c} + r^2} 
        \\
        \Lambda u_{r \rm c} 
        \\
        u_{z \rm c} 
        \end{pmatrix},
    \\
    & e_\mu^{3} = - \frac{1}{\sqrt{K_{\rm c}}}
        \begin{pmatrix}
        - \frac{a}{\Sigma} \left(r (1 - z^2) u_{z \rm c} + z \Delta u_{r \rm c}\right) 
        \\
        \frac{1 - z^2}{\Sigma}\left(r (r^2 + a^2) u_{z \rm c} + a^2 z \Delta u_{r \rm c} \right)
        \\
        \frac{az}{\Delta}\left( (r^2 + a^2)E_{\rm c} -a L_{\rm c} \right)
        \\
        a r E_{\rm c} - \frac{r L_{\rm c}}{1 - z^2} 
        \end{pmatrix}, \\
    & \Lambda \equiv  \frac{K_{\rm c} - a^2 z^2}{K_{\rm c} + r^2}\,,
\end{align}
\end{subequations}
where we used the $t,\phi,r,z$ ordering of components with respect to Boyer-Lindquist coordinates (we further use $z= \cos \theta$ as a coordinate instead of the polar angle $\theta$ to obtain expressions in terms of rational functions). The constants $E_{\rm c}, L_{\rm c}, K_{\rm c}$ represent specific energy, azimuthal angular momentum, and Carter constant \cite{Carter:1968rr}, all of which are constant throughout the congruence. The $u_{r \rm c}, u_{z \rm c}$ components are given by
\begin{align}
    & u_{r \rm c} = \pm \frac{\sqrt{R(r;E_{\rm c}, L_{\rm c},K_{\rm c})}}{\Delta} \,, \label{eq:urc}
    \\
    & u_{z \rm c} = \pm \frac{\sqrt{Z(z;E_{\rm c}, L_{\rm c},K_{\rm c})}}{1 - z^2}\,, \label{eq:uzc}
    \\
    & R(r;E,L,K) \equiv \left(E(r^2 +a^2) - a L\right)^2 - \Delta( K + r^2) \,, \label{eq:Rconst}
    \\
    & Z(z; E,L,K) \equiv (1-z^2)\left(K - a^2 z^2 \right) - \left(L - a E(1-z^2) \right)^2 \!.\! \label{eq:Zconst}
\end{align}
The constants $E_{\rm c}, L_{\rm c}, K_{\rm c}$ determine the geodesic congruence and thus also the tetrad field almost uniquely up to the choice of the direction of the $r,z$ motion reflected in the sign choices in the expressions for $u_{r \rm c}, u_{z \rm c}$. Using the information above, it is straightforward to compute the Ricci rotation components ${e}_{A;\mu}^\kappa  {e}_{B \kappa}$ (the interested reader can find the explicit expressions in Appendix C of Ref. \cite{Piovano:2024yks}). The Marck tetrad congruence was then used in Ref. \cite{Witzany:2019nml} to generate the canonical coordinates on the spinning particle phase space and to formulate the Hamilton-Jacobi equation.

 The tetrad field is real only within a ``box'' of turning points for bound motion, $r\in(r_2,r_1),\, z \in (- z_2, z_2)$. The canonical coordinates for the spinning particle should also be seen as defined only within this coordinate box. The turning points are a subset of the roots of $R,Z$. Following \citet{Schmidt:2002qk}, one can always pick the physical turning points first, and from this determine the other two roots of $R$ denoted as $r_3$ and $r_4$, and the larger (degenerate) root of $Z$, denoted as $z_{1}$ as well as the constants of motion. 
 The functions $r_3(r_1,r_2,z_2),\, r_4(r_1,r_2,z_2),\, z_1(r_1,r_2,z_2)$ 
 and $E(r_1,r_2,z_2),\, L(r_1,r_2,z_2),\, K(r_1,r_2,z_2)$ are complicated but explicit, and can be found in the Appendix of Ref. \cite{Schmidt:2002qk} (they are also implemented in the KerrGeodesics Mathematica package \cite{KerrGeodesics090}). The functions $R,Z$ are then easily re-expressed in factorized form as
\begin{align}
    & R(r;r_1,r_2,z_2) = (1 - E^2) (r_1 - r)(r-r_2)(r-r_3)(r-r_4) \,, \label{eq:Rroot}
    \\
    & Z(z;r_1,r_2,z_2) = a^2(1-E^2) (z_1^2 - z^2)(z_2^2 - z^2)\,, \label{eq:Zroot}
\end{align}
where it is assumed that we substitute $E = E(r_1,r_2,z_2)$.

The Marck tetrad \eqref{eq:mtet} allows for the separation of parallel transport along geodesics and the solution can consequently be written in terms of elliptic integrals, as noted in the original paper by Marck \cite{marck1983solution} (a convenient summary of Marck's work and a reduction of these integrals into Legendre canonical form was given by \citet{vandeMeent:2019cam}). However, it is important to note that the connection $e_{A;\mu}^\kappa  {e}_{B \kappa}$ determined by the Marck tetrad congruence is singular near the turning points since $u_{y \rm c}$ goes to zero as $\sim \sqrt{y - y_{ \rm tp}}$ for $y = r,z$ and $y_{\rm tp}$ the turning point. This can be seen by comparing equations \eqref{eq:urc} and \eqref{eq:Rroot} for $u_{r \rm c}$ and equations \eqref{eq:uzc} and \eqref{eq:Zroot} for $u_{z \rm c}$.
As such, the derivatives of $u_{y \rm c}, y=r,z,$ with respect to the space-time coordinates
appearing in the connection diverge as $\sim 1/\sqrt{y - y_{\rm tp}}$ near turning points. This must be taken into account when counting orders within the scheme.

\subsection{Solution of the Hamilton-Jacobi equation} \label{subsec:hamjac}
Using the canonical coordinates generated by the Marck tetrad one can perturbatively solve the Hamilton-Jacobi equation for the spinning particle in Kerr space-time \cite{Witzany:2019nml}. The Tulczyjew-Dixon condition reduces the spin sector to one relevant degree of freedom, characterized by a canonical momentum $\pi_\psi$ and a conjugate coordinate $\psi \in [0,2\pi)$ (see Refs \cite{Witzany:2018ahb,Witzany:2019nml} for details). The resulting Hamilton's principal function (``the action'') in canonical coordinates $t,\phi,r,z,\psi$ is given by 
\begin{align}
\label{eq:W}
\begin{split}
    &W = 
        -E_{\rm so} (t - t_0) + L_{\rm so} (\phi - \phi_0) + (s_\parallel - s) (\psi - \psi_0)  
        \\
        &\phantom{W = }
        + \!\sum_{y = r,z} \int\! \pm \left(\!
            \sqrt{w_y'^2 + e_{0y}{e}_{C;y}^\kappa  {e}_{D \kappa} \tilde{s}^{CD}} - \frac{1}{2} {e}_{A;\mu}^\kappa  {e}_{B \kappa} \tilde{s}^{AB} 
        \!\right)\! \mathrm{d} y, 
\end{split}
\end{align}
where the auxiliary functions $w_y'(y)$, $y=r,z$ are defined as (primes denote derivatives with respect to the function arguments)
\begin{align}
\begin{split}
    & \left(\Delta \frac{\mathrm{d} w_r}{\mathrm{d} r}\right)^2 \equiv 
        -\Delta (K_{\rm so} + r^2) + [(r^2+a^2)E_{\rm so} - a L_{\rm so}]^2
        \\
        & \phantom{\left(\Delta \frac{\mathrm{d} w_r}{\mathrm{d} r}\right)^2 \equiv}
        +2  s_\parallel \Delta \sqrt{K_\mathrm{c}} \frac{ E_\mathrm{c} (r^2 + a^2) - a L_\mathrm{c}}{K_\mathrm{c}+ r^2}   \,,
\end{split} \label{eq:wrp}
    \\
\begin{split}
    & \left((1 -z^2) \frac{\mathrm{d}w_z}{\mathrm{d}z}\right)^2 \equiv 
        (1-z^2)\left(K_{\rm so} - a^2 z^2 \right)
        \\ 
        & \phantom{(1 -z^2) \frac{\mathrm{d}w_z}{\mathrm{d}z}^2 \equiv }
        - \left(L_{\rm so} - a E_{\rm so}(1-z^2) \right)^2 
        \\ 
        & \phantom{(1 -z^2) \frac{\mathrm{d}w_z}{\mathrm{d}z}^2 \equiv }
        + 2 a s_\parallel (1 - z^2) \sqrt{K_{\rm c}} \frac{L_{\rm c} - a E_{\rm c} (1 - z^2)}{K_{\rm c} - a^2 z^2}  \,.
\end{split} \label{eq:wzp}
\end{align}
Additionally, we used the leading-order spin tensor 
\begin{align}
    \tilde{s}^{0A} = & 0\,, \\
    \tilde{s}^{12} = & - \tilde{s}^{21} =  s_\parallel \,,\\
    \tilde{s}^{23} = & - \tilde{s}^{32} =  \sqrt{s^2 - s_\parallel^2} \cos \psi \,,\\
    \tilde{s}^{31} = & -\tilde{s}^{13} = \sqrt{s^2 - s_\parallel^2} \sin \psi \,.
\end{align}
The tensor $\tilde{s}^{AB}$ agrees with $s^{AB}$ up to $\mathcal{O}(s^{3/2})$ terms near turning points \cite{Witzany:2019nml}.
The constants $E_{\rm so},L_{\rm so}, K_{\rm so}, s_\parallel$ in eq. \eqref{eq:W} are separation constants and $\mathcal{O}(s)$ constants of motion. The arbitrary $t_0,\phi_0,\psi_0$ are integration constants (similar $r_0,z_0$ constants are implicit in the indefinite integrals in eq. \eqref{eq:W}). 

$E_{\rm so}$ and $L_{\rm so}$ are the specific spin-orbital energy and azimuthal angular momentum generated by Killing symmetries \cite{Dixon:1970I}. $K_{\rm so}$ is a spin deformation of the Carter constant identical to the quadratic constant found by \citet{Rudiger:1983}, and $s_\parallel<s$ denotes the spin component aligned with the orbital angular-momentum vector; it is proportional to the linear constant found by \citet{Rudiger:1981} (for more, see Ref. \cite{Witzany:2019nml}). The $\pm$ sign choices correspond to the switching of the direction of the $r,z$ motion.

\begin{figure*}
    \centering
    \includegraphics[width=0.8\linewidth]{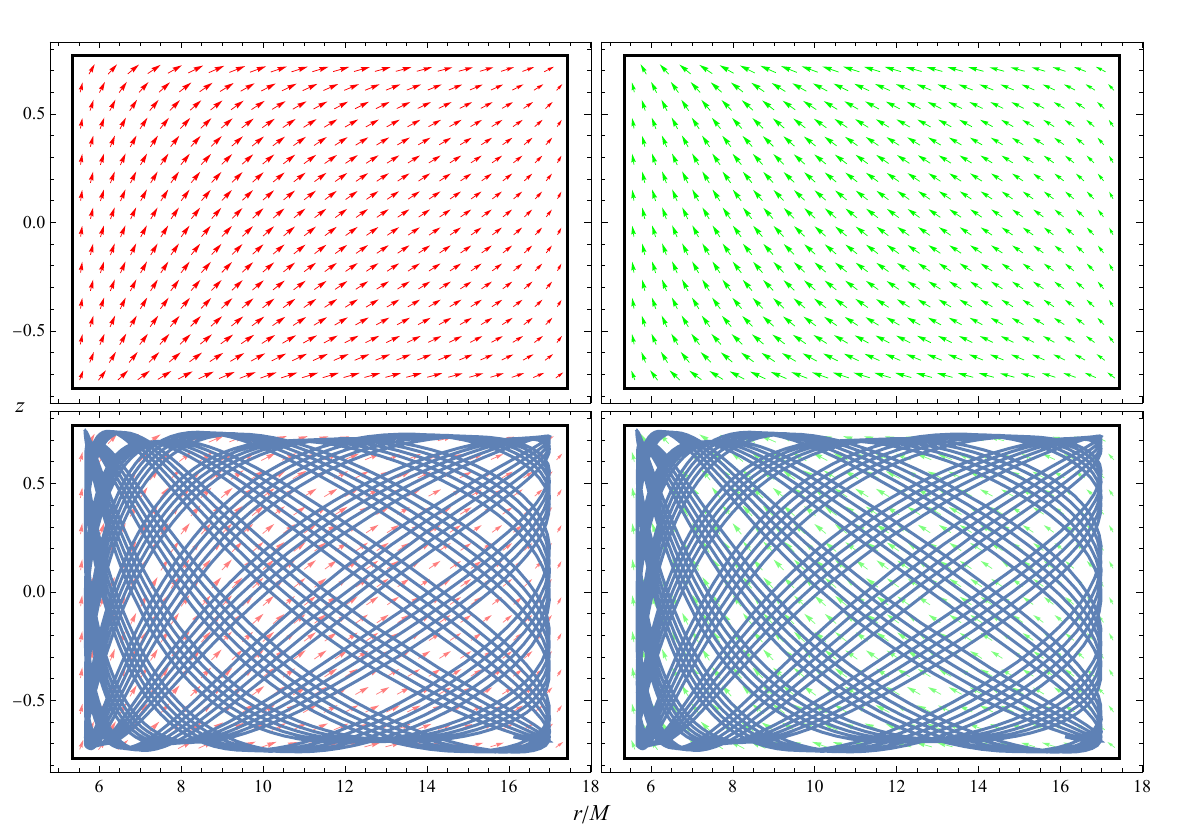} 
    \caption{Top row: The $r$-$z$ projection of the $e^{0}_\mu$ leg of the Marck tetrad congruence with $E_{\rm c} = 0.959, L_{\rm c} = 2.14, K_{\rm c} = 8.29$ and different branch choices of $u_{r \rm c}$. Bottom row: Spinning particle trajectory with $E_{\rm so} = 0.958, L_{\rm so} = 2.39, K_{\rm so} = 8.37$ and $s_{\parallel} = 0.1 M, s = 0.141 M$ superposed on top of the congruence (the value of secondary spin is exaggerated to observe the effects more clearly). All plots are for primary black hole spin $a = 0.9M$. Plotted in thick black in all plots is also the box of turning points of the background congruence. The derivatives of $e^A_{\mu}$ and thus the Ricci rotation coefficients diverge along the box. However, the $O(s)$ freedom in picking the congruence constants allowed us to avoid hitting this singularity along the particle orbit.}
    \label{fig:CongruenceIllustration}
\end{figure*}

The variables $E_{\rm so}, L_{\rm so}, K_{\rm so}$ are assumed to be $\mathcal{O}(s)$ close to $E_{\rm c}, L_{\rm c}, K_{\rm c}$ defining the background tetrad congruence. Even more specifically, one has to assume that the constants $E_{\rm c}, L_{\rm c}, K_{\rm c}$ are chosen so that the ``geodesic box'' $r\in (r_1,r_2), z\in(-z_2,z_2)$ of the congruence encloses the motion of the spinning particle with an $\mathcal{O}(s)$ distance of the boundaries of the geodesic box from the spinning particle. As a result, the connection terms reach $\sim 1/\sqrt{s}$ order near the turning points, but at least do not diverge. This is an important point for the mathematical well-definedness of the procedure, which we illustrate in Fig. \ref{fig:CongruenceIllustration}.

The gradient of $W$ with respect to any of the canonical coordinates $t,\phi,r,z,\psi$ gives the value of its canonically conjugate momentum. As discussed in Ref. \cite{Witzany:2019nml}, the momenta $W_{,y} = \pi_{y},\, y=r,z$ have $\sim 1\sqrt{s}$ singularities near turning points. This can be verified by computing the $\partial/\partial y$ derivative of Eq. \eqref{eq:W} and noticing the appearance of the near-singular connection terms discussed in the previous paragraph.

These $1/\sqrt{s}$ near-singularities in $\pi_y$ would seem to limit the physical validity of the solution. 
However, the equations of motion following from the Hamiltonian \eqref{eq:HTD} yield
\begin{align}
\label{eq:EOM}
\begin{split}
     \frac{{\rm d} y}{ \rm d \tau} 
        & = g^{yy} \left(\pi_y + \frac{1}{2} {e}_{A;y}^\kappa  {e}_{B \kappa} \tilde{s}^{AB} \right) 
        \\
        & = \pm g^{yy} \sqrt{w_y'^2 + e_{0y}{e}_{C;y}^\kappa  {e}_{D \kappa} \tilde{s}^{CD}} + \mathcal{O}(s^{3/2})\,,
\end{split}
\end{align}
which are smooth regular functions at $\mathcal{O}(s)$. The $\sim 1/\sqrt{s}$ singularities can thus be understood simply as coordinate artefacts that are consistently eliminated from any physical observable. On the other hand, we see that the various connection terms in $W$ \textit{cannot} be summarily discarded and are important to obtain the correct final form of the local equations of motion. At the same time, the connection terms are not separable, and the motion is thus \textit{not} separable in Boyer-Lindquist coordinates. 

The correctness of the non-separable equations of motion \eqref{eq:EOM} and the practical usefulness of this formalism was recently directly confirmed by \citet{Piovano:2024yks} by comparing the semi-analytical integration of eq. \eqref{eq:EOM} against the spectral integration of the Mathisson-Papapetrou-Dixon equations by \citet{Drummond:2022efc,Drummond:2022xej}.

\subsection{Defining actions} \label{subsec:defactions}

Actions are invariants defined by the geometry of an integrable dynamical system (see, e.g., the textbook of \citet{Arnold:1989who} and the monograph of \citet{Arnold:2006}). To define them, we use the Poincaré-Cartan form $\sum_{q}\pi_{q} {\rm d}q$ where $q= t,\phi,r,z,\psi$ are the canonical coordinates. Next, we take hypersurfaces of constant $E_{\rm so},L_{\rm so}, K_{\rm so}, s_\parallel$ and express $\pi_q$ as a function of $t,\phi,r,z,\psi$ on these hypersurfaces. The hypersurfaces of constant integrals of motion usually have a nontrivial topology, and the expression of $\pi_q$ on them thus do not need to be single-valued when expressed as functions of $q$. Such ambiguities arise because the motion consists of librations that return with different momentum signs, as in the case of $r,z$ motion. 

Given a loop $\gamma$ over the constant-integral hypersurface, the actions are defined as
\begin{align}
    J_\gamma = \frac{1}{2\pi} \oint_\gamma \sum_q \pi_q \mathrm{d}q\,.
\end{align}
The actions are in fact defined only by the homotopy class of $\gamma$, since any smooth deformation of $\gamma$ leads to an action that differs only by an integral of $\sum_q \mathrm{d}\pi_q\wedge\mathrm{d}q$ over a contractible curve, which is zero (see Refs \cite{Arnold:1989who,Arnold:2006,fasano2006analytical} for more).  In this sense, the actions are \textit{gauge-invariant}, since they do not depend on the choice of phase-space coordinates or small deformations thereof.

This invariance implies that once a set of actions is defined, any other set can be derived from it using a discrete lattice-type transform with a determinant equal to $\pm1$.\footnote{The determinant condition excludes cases of curves with multiple winding around the torus. This is because actions defined by multiply wound curves would be canonically conjugate to angles that are multiply valued on the invariant tori in phase space  \cite{fasano2006analytical}.} Notably, this transformation can involve discrete jumps in its coefficients rather than being smooth. 

If the constant-integral hypersurface is non-compact in some direction, such as $t$, additional action variables can be chosen for integration over infinite curves on the hypersurfaces. However, the normalization of these additional action variables is non-unique and must be determined by external physical arguments.  

We choose the loops as follows
\begin{align}
    & \gamma_{r}:r \text{ loop};\,t,\phi,z,\psi \text{ const.};\, z\text{ far from turning points,} \\
    & \gamma_{z}:z\text{ loop};\,t,\phi,r,\psi \text{ const.};\, r\text{ far from turning points,} \\  
    &\gamma_{\phi}\!:\phi\text{ loop};\,t,r,z,\psi \text{ const.};\, r,z\text{ far from turning points,} \\
    &\gamma_{\psi}\!:\psi\text{ loop};\,t,r,z,\phi \text{ const.};\, r,z\text{ far from turning points.} 
\end{align}
These define actions that we denote as $J_y (y=r,z),J_\phi, J_\psi$. At $\mathcal{O}(s)$ they reduce to
\begin{align}
    & J_y = \frac{1}{2\pi}\oint_{\gamma_{y}} \!\pm\left(\!
            \sqrt{w_y'^2 + e_{0y}{e}_{C;y}^\kappa  {e}_{D \kappa} \tilde{s}^{CD}} - \frac{1}{2} {e}_{A;y}^\kappa  {e}_{B \kappa} \tilde{s}^{AB} 
        \!\right)\! \mathrm{d} y \,, \label{eq:J_y_def}\\
    & J_\phi = L_{\rm so} \,,\\
    & J_\psi = s_\parallel -s \,.
\end{align}
For completeness, we define the action $J_t$ by the infinite curve $\gamma_t$ spanning $t\in \mathbb{R}$ and $\phi,r,z,\psi$ arbitrary but $r,z$ away from the turning points. Normalization is set by unit coordinate time or
\begin{align}
    J_t \equiv \lim_{T \to \infty} \frac{1}{T} \int_{-T/2}^{T/2} \pi_t \mathrm{d} t  = -E_{\rm so} \,.
\end{align}
This choice is motivated by $t$ being the time of inertial observers at infinity.  

\subsection{Computing actions} \label{subsec:compactions}
The issue with evaluating the expressions for $J_y$ in equation \eqref{eq:J_y_def} is that the connection terms depend on both $r, z$, while the turning points are functions of the second coordinate ($z$ for $r$ turning points and vice versa). This complicates the expansion in spin $s$. A curious property is also the fact that the $\gamma_r$ loop and thus the $J_r$ action seem to be a function of the $z$ coordinate that is held constant and vice versa for $\gamma_z$ and $J_z$. However, the resulting expressions for $J_r,J_z$ should be insensitive to the values of $z,r$ that are held constant along the $\gamma_{r,z}$ loops as long as the homotopy class of the phase-space curve is conserved when choosing a different fixed $r,z$. 

In fact, it was shown by \citet{Damour:1988mr} that loop integrals appearing in actions can be viewed as contour integrals in the complex plane which are ``topological''. One can then obtain an expansion of such integrals in a perturbative parameter by the following procedure. First, expand the integrand while ignoring the dependence of the integration bounds on the perturbative parameter. Some of the resulting terms in that expansion may be singular, which is resolved by regularizing them by Hadamard's partie finie analytical continuation (see Appendix B of Ref. \cite{Damour:1988mr}). The result is then equivalent to the expression obtained by the expansion of an exact evaluation of the integral. 

We use this method in the case of the $J_y$ actions and the small secondary spin parameters $s_\parallel, s$. We start by partially expanding $J_y$ as
\begin{align}
\begin{split}
    & J_y = \oint_{\gamma_{y}} \!\pm\left(\!
            \sqrt{w_y'^2 + e_{0y}{e}_{C;y}^\kappa  {e}_{D \kappa} \tilde{s}^{CD}} - \frac{1}{2} {e}_{A;y}^\kappa  {e}_{B \kappa} \tilde{s}^{AB} 
        \!\right)\! \mathrm{d} y \,, \\
    & \phantom{J_y} = \frac{1}{\pi} {\rm Pf} \int_{y_{\rm g tp}}  \sqrt{w_y'^2} + \frac{1}{2} \left(\frac{e_{0y}}{\sqrt{w_y'^2}} -1 \right) {e}_{C;y}^\kappa  {e}_{D \kappa} \tilde{s}^{CD} \mathrm{d} y 
    \\ 
    & \phantom{J_y =} + \mathcal{O}(s^2)\,,
\end{split}
\end{align}
where $\rm Pf$ denotes partie finie integration, and $y_{\rm gtp}$ are turning points of a geodesic close to the spinning particle. 
Here the partie finie integration regularizes any of the divergences that may occur at $\mathcal{O}(s^2)$ around turning points and we can thus keep a naive order counting of spin behind the integral sign. Next we notice from \eqref{eq:wrp} and \eqref{eq:wzp} and from the Marck congruence construction from Section \ref{subsec:marck} that we have $e_{0 y} = \sqrt{w_y'^2} + \mathcal{O}(s)$ up to turning point singularities. As a result, the connection terms cancel out at $\mathcal{O}(s)$. We are then left only with the fully separable integrand $\sqrt{w_y'^2}$ up to $\mathcal{O}(s^2)$. Finally, we use the expressions for $w_y'^2$ given in eq. \eqref{eq:wrp} and \eqref{eq:wzp} to expand the integrals as $J_y = J_{y}^{(0)} + J_{y}^{(1)}$ where 
\begin{align}
      \label{eq:J_r_0_def}
  & J_{r}^{(0)}  \equiv \frac{1}{2\pi} \oint \frac{\pm \sqrt{R_{\rm so}(r)}}{\Delta} \mathrm{d}r
  = \frac{1}{\pi} \int_{r_{2}}^{r_{1}} \frac{\sqrt{R_{\rm so}(r)}}{\Delta} \mathrm{d}r \,,\\
\begin{split}
  & J_{r}^{(1)}  \equiv \frac{s_{\parallel}}{2\pi} \oint \! \frac{\pm \Delta}{2\sqrt{R_{\rm so}(r)}}
  \left[ \!
    2  \sqrt{K_\mathrm{c}} \frac{ E_\mathrm{c} (r^2 + a^2) - a L_\mathrm{c}}{\Delta(K_\mathrm{c}+ r^2)}
  \right] \mathrm{d}r
  \,,  \\
  \label{eq:J_r_1_def}
  & \phantom{J_{r}^{(1)}}  = \frac{s_{\parallel}}{\pi} \int_{r_{2}}^{r_{1}} \! \frac{1}{\sqrt{R_{\rm so}(r)}}
  \left[
    \sqrt{K_\mathrm{c}} \frac{ E_\mathrm{c} (r^2 + a^2) - a L_\mathrm{c}}{K_\mathrm{c}+ r^2}
  \right] \mathrm{d}r
  \,,
\end{split}
  \\
  \label{eq:Rso}
  & R_{\rm so}(r) \equiv R(r; K_{\rm so},E_{\rm so}, L_{\rm so}) \,,
\end{align}
and 
\begin{align}
  \label{eq:J_z_0_def}
  & J_{z}^{(0)} \equiv \frac{1}{2\pi} \oint \frac{\pm \sqrt{Z_{\rm so}(z)}}{1 - z^2} {\rm d}z
  = \frac{2}{\pi} \int_0^{z_2} \frac{\sqrt{Z_{\rm so}(z)}}{1 - z^2} {\rm d}z \,,\\
\begin{split}
  & J_{z}^{(1)}  \equiv \frac{s_{\parallel}}{2\pi} \oint \frac{\pm 1}{2\sqrt{Z_{\rm so}(z)}}
  \left[
    2 a \sqrt{K_{\rm c}} \frac{L_{\rm c} - a E_{\rm c} (1- z^2)}{K_{\rm c} - a^2 z^2}
  \right] \mathrm{d} z
  \,,  \\
  \label{eq:J_z_1_def}
  &\phantom{J_{z}^{(1)}} = \frac{2s_{\parallel}}{\pi} \int_{0}^{z_2}\!\! \frac{1}{\sqrt{Z_{\rm so}(z)}}
  \left[
    a \sqrt{K_{\rm c}} \frac{L_{\rm c} - a E_{\rm c} (1- z^2)}{K_{\rm c} - a^2 z^2}
  \right] \mathrm{d}z
  \,, 
\end{split}
  \\
  & Z_{\rm so}(z) \equiv Z(z;E_{\rm so},L_{\rm so},K_{\rm so})\,.
\end{align}
In other words, we can express the actions as sums of geodesic actions of a referential geodesic with $E= E_{\rm so}, L=L_{\rm so}, K=K_{\rm so}$ and spin corrections. Note that the corresponding roots $r_{1,2}, z_2$ are not true turning points of the spinning particle but of this referential geodesic and, furthermore, that the choice of the referential geodesic is not unique.\footnote{Additionally, note that these ``virtual'' or referential geodesics in the integrals are generally only $\mathcal{O}(s)$ close to the background ``congruence'' geodesics generating the tetrad $e^A_\mu$. This distinction is important for the well-definedness of the construction, but in practice one can set $C_{\rm c} = C_{\rm so},$ for all $C=E,L,K$, since the difference only leads to $\mathcal{O}(s^2)$ corrections.} As such, one must keep in mind that only the total sums $J_y^{(0)} + J_y^{(1)}$ are physically meaningful.

Fortuitously, none of the expanded expressions are divergent here, so no Partie finie regularization is needed at this point. The integrands of $J_r^{(0,1)}, J_{z}^{(0,1)}$ can be expressed as a square root of a polynomial up to fourth order multiplied by a rational function. As such, the integrals are elliptic integrals that can be expressed in Legendre canonical form \cite{byrd2013handbook}. Additionally, the integration runs between roots of the polynomial under the square root, which means that all the involved integrals will be complete. 

We carried out the reduction of the integrals to Legendre form with the resulting expressions for the radial part:
\begin{widetext}
\begin{align}
  \begin{split}
  \label{eq:J_r_0_eval}
    & J_{r}^{(0)} =
    \frac{(r_{2}-r_{3})\sqrt{1-E_{\rm so}^{2}}}
    {\pi\sqrt{(r_{1}-r_{3})(r_{2}-r_{4})}} \Bigg[ 
        (r_{1}-r_{3})\mathsf{K}(k_{r}) -
        \frac{(r_{1}-r_{3})(r_{2}-r_{4})}{r_{2}-r_{3}} \mathsf{E}(k_{r})
        +
        \frac{2(r_{1}-r_{+})(r_{4}-r_{+})}{r_{+}-r_{-}}
        \mathsf{\Pi}(\rho_{+}^{-2}, k_{r})
        \\
        &\phantom{J_{r}^{(0)} =}
        -
        \frac{2(r_{1}-r_{-})(r_{4}-r_{-})}{r_{+}-r_{-}}
        \mathsf{\Pi}(\rho_{-}^{-2}, k_{r})
        - \frac{(1-2 E_{\rm so}^{2})(r_{+}+r_{-})}{1-E_{\rm so}^{2}}
        \mathsf{\Pi}(\alpha_{r}^{2},k_{r})
    \Bigg] \,,
  \end{split}
    \\
    \label{eq:J_r_1_eval}
    &J_{r}^{(1)} =
        \frac{s_{\parallel}}{\pi}
        \frac{2 \sqrt{K_{\rm so}} (a^2 E_{\rm so} - a L_{\rm so} + E_{\rm so} r_{3}^2)}
        {\sqrt{1 - E_{\rm so}^2} (K_{\rm so} + r_{3}^2) \sqrt{(r_{1} - r_{3}) (r_{2} - r_{4})}}
        \left[
            \mathsf{K}(k_{r})
            - \frac{\gamma_{r,1,+}}{\rho_{p,+}^{2}}\mathsf{\Pi}(\rho_{p,+}^{-2},k_{r})
            - \frac{\gamma_{r,1,-}}{\rho_{p,-}^{2}}\mathsf{\Pi}(\rho_{p,-}^{-2},k_{r})
        \right]
    \,, \\
\end{align}
where
\begin{align}
    \label{eq:k_r_def}
    & k_{r}^{2} \equiv \frac{(r_{1} - r_{2}) (r_{3} - r_{4})}{(r_{1} - r_{3}) (r_{2} - r_{4})}
    \,, \quad 
    r_{\pm}\equiv M \left(1 \pm \sqrt{1-a^{2}/M^{2}} \right)
    \,, \quad
    \rho_{\pm}^{2} \equiv \frac{(r_{\pm}-r_{2})(r_{1}-r_{3})}{(r_{1}-r_{2})(r_{\pm}-r_{3})}
    \,, \quad 
    \alpha_{r}^{2} = \frac{r_{1}-r_{2}}{r_{1}-r_{3}}
    \,. \\
    & \label{eq:rho_pz_def}
    \rho_{z,\pm}^{2} \equiv 
    \frac{r_{1} - r_{3}}{r_{1} - r_{2}} \times
    \frac{a^2 E_{\rm so} - a L_{\rm so} + E_{\rm so} r_{2} r_{3} \pm (r_{2} - r_{3})\sqrt{a E_{\rm so} (- a E_{\rm so} + L_{\rm so})} }
    {a^2 E_{\rm so} - a L_{\rm so} + E_{\rm so} r_{3}^2}
    \,,\quad
    \rho_{p,\pm}^{2} \equiv
    \frac{r_{1}-r_{3}}{r_{1}-r_{2}}
    \frac{K_{\rm so} + r_{2} r_{3} \pm (r_{2}-r_{3}) \sqrt{- K_{\rm so}} }
    {K_{\rm so} + r_{3}^2}
    \,, \\ 
    & \label{eq:gamma_r1_def}
    \gamma_{r,1,\pm} \equiv \pm \frac{(\rho_{p,\pm}^{2}-\rho_{z,+}^{2})(\rho_{p,\pm}^{2}-\rho_{z,-}^{2})}{\rho_{p,+}^{2}-\rho_{p,-}^{2}}
    = \pm \frac{(r_{1} - r_{3}) (r_{2} - r_{3})(a^2 E_{\rm so} - E_{\rm so} K_{\rm so} - a L_{\rm so}) (- K_{\rm so} + r_{3}^{2} \pm 2 r_{3}\sqrt{- K_{\rm so}})}
    {2 (r_{1} - r_{2}) \sqrt{- K_{\rm so}} (K_{\rm so} + r_{3}^2) (a^2 E_{\rm so} - a L_{\rm so} + E_{\rm so} r_{3}^2)}\,.
\end{align}
\end{widetext}
 For the polar part we obtained the much simpler expressions 
\begin{align}
\begin{split}
\label{eq:J_z_0_eval}
    & J_{z}^{(0)} =
      \frac{2a\sqrt{1-E_{\rm so}^{2}}}{\pi}
      \Big[
        - \frac{1-z_2^{2}}{z_1}\mathsf{K}\left(\frac{z_2 }{z_1}\right)
        + z_1 \mathsf{E}\left(\frac{z_2 }{z_1}\right)
        \\ 
        &\phantom{ J_{z}^{(0)} = \frac{2a\sqrt{1-E_{\rm so}^{2}}}{\pi} \Big[} 
         + \frac{(1-z_2^{2})(1-z_1^{2})}{z_1} \mathsf{\Pi}\left(z_2^{2},\frac{z_2}{z_1}\right)
      \Big] \,,
\end{split}
\\
\begin{split}
\label{eq:J_z_1_eval}
    & J_{z}^{(1)} =
   \frac{2 s_{\parallel}}{\pi} 
  \frac{1}{a z_1\sqrt{K_{\rm so}}\sqrt{1-E_{\rm so}^{2}}} \times
   \\
  & \phantom{J_{z}^{(1)} =\,} \Big[
    (E_{\rm so}(K_{\rm so}-a^2)+aL_{\rm so})\mathsf{\Pi}\left(\frac{a^{2}z_2^{2}}{K_{\rm so}},\frac{z_2}{z_1}\right)
    \\
    & \phantom{J_{z}^{(1)} =\, \Big[} 
    - E_{\rm so}K_{\rm so} \ \mathsf{K}\left(\frac{z_2}{z_1}\right)
  \Big]
  \,.
\end{split}
\end{align}
The $\mathsf{K}(k), \mathsf{E}(k), \mathsf{\Pi}(k)$
are complete elliptic integrals of the first, second, and third kind. We follow the conventions of the \textit{Digital Library of Mathematical Functions} in their definition~\cite{NIST:DLMF}, see also Appendix \ref{app:leg}. The detailed steps leading to this result are also given in Appendix \ref{app:leg}.

\subsection{Fundamental frequencies from actions} \label{subsec:frequencies}

Since we obtained closed-form expressions for the actions, we can compute fundamental frequencies of motion by using the implicit method introduced by Schmidt for Kerr geodesics~\cite{Schmidt:2002qk}. This is very useful, since these frequencies are a necessary ingredient for the frequency-domain computation of gravitational waves emergent from the motion of the spinning particle, and previous computations had to evaluate these frequencies numerically or semi-analytically \cite{Drummond:2022xej,Drummond:2022efc,Skoupy:2023lih,Piovano:2024yks}. 

Schmidt's technique exploits the fact that even though we do not know closed functional forms of action-angle Hamiltonians generating motion parametrized by various choices of time parameters, we can still compute their derivatives through the use of the implicit function theorem.  

We start by computing partial derivatives
\begin{align}
  \frac{\partial J_{k}}{\partial C_{i}}\Big|_{\text{other }C\text{ const.}} = \left(\frac{\partial J^{(0)}_{k}}{\partial C_{i}} + \frac{\partial J^{(1)}_{k}}{\partial C_{i}}\right)\Big|_{\text{other }C\text{ const.}},
\end{align}
for the actions labeled by $k$ with respect to the
constants of motion
$\vec{C}=(H_\tau, K_{\rm so}, E_{\rm so}, L_{\rm so}, s_{\parallel})$, where $H_\tau = H_\text{TD}$ is the Hamiltonian generating evolution in proper time. Both $\vec{C}$ and
$\vec{J} = (J_{r}, J_{\theta}, J_{t}, J_{\phi}, J_{s})$ are assumed to be independent coordinates parametrizing the integral hypersurfaces of the motion, and as such there is an invertible
transformation between them. We can then compute the Jacobian of the inverse transform simply as a matrix inverse or
\begin{align}
  \frac{\partial C_{i}}{\partial J_{k}}\Big|_{\text{other }J\text{ const.}} = \left[ \frac{\partial J}{\partial C}\Big|_{\text{other }C\text{ const.}} \right]^{-1}_{ik}
  \,.
\end{align}
Within the matrix
$\partial C_{i}/\partial J_{k}$, the derivatives of $H_\tau = -1/2$ with respect to the actions generate proper-time frequencies $\omega^k$.
The elements in that row are
\begin{subequations}\label{eq:frequencies_Schmidt}
\begin{align}
  \omega^{r} &= \frac{\partial H_\tau}{\partial J_{r}} = \frac{-1}{\mathcal{D}} \frac{\partial J_{z}}{\partial K_{\rm so}} \,,\\
  \omega^{z} &= \frac{\partial H_\tau}{\partial J_{z}} = \frac{+1}{\mathcal{D}} \frac{\partial J_{r}}{\partial K_{\rm so}} \,,\\
  \omega^{t} &= \frac{\partial H_\tau}{\partial J_{t}} = \frac{1}{\mathcal{D}}
  \left[ \frac{\partial J_{r}}{\partial K_{\rm so}} \frac{\partial J_{z}}{\partial E_{\rm so}} - \frac{\partial J_{r}}{\partial E_{\rm so}} \frac{\partial J_{z}}{\partial K_{\rm so}} \right]
  \,,\\
  \omega^{\phi} &= \frac{\partial H_\tau}{\partial J_{\phi}} = \frac{1}{\mathcal{D}}
  \left[ \frac{\partial J_{r}}{\partial L_{\rm so}} \frac{\partial J_{z}}{\partial K_{\rm so}} - \frac{\partial J_{r}}{\partial K_{\rm so}} \frac{\partial J_{z}}{\partial L_{\rm so}} \right]
  \,,\\
  \omega^{\psi} &= \frac{\partial H_\tau}{\partial J_{\psi}} = \frac{1}{\mathcal{D}}
  \left[ \frac{\partial J_{r}}{\partial s_{\parallel}} \frac{\partial J_{z}}{\partial K_{\rm so}} - \frac{\partial J_{r}}{\partial K_{\rm so}} \frac{\partial J_{z}}{\partial s_{\parallel}} \right]
  \,,\\
  \mathcal{D} &= \frac{\partial J_{r}}{\partial K_{\rm so}} \frac{\partial J_{z}}{\partial H_\tau} - \frac{\partial J_{r}}{\partial H_\tau} \frac{\partial J_{z}}{\partial K_{\rm so}}
  \,.
\end{align}
\end{subequations}
The frequencies with respect to the coordinate time can be calculated as
\begin{equation}
    \Omega^k = \frac{\omega^k}{\omega^t}\,,
\end{equation}
where $k = r,z,\phi,\psi$.

There is an issue when computing the derivatives $\partial J_y/\partial H_{\tau}$ and that is that we have substituted the strictly on-shell value of the Hamiltonian $H_\tau = -1/2$ during the derivation of $J_y$ and did not keep the value of $H_\tau$ as a parameter of the action. As a result, we are unable to directly compute the derivatives. To amend this, we would have to retrace the derivation of the actions with the off-shell value of $H_\tau$. However, in the final expressions for $\Omega^k$, the determinant 
$\mathcal{D}$
of the matrix $\partial \vec{J}/\partial \vec{C}$, which contains the derivatives with respect to the Hamiltonian $H_\tau$, cancels out. Therefore, these derivatives are not needed for the coordinate-time frequencies $\Omega^k$.

Since $J_r$ and $J_z$ are expressed as algebraic functions of elliptic integrals, so will their derivatives. However, calculating derivatives of the expressions in Eqs.~\eqref{eq:J_r_0_eval} -- \eqref{eq:J_z_1_eval} would be too laborious and the resulting expressions would be too long. Instead, we can take derivatives of the defining integrals \eqref{eq:J_r_0_def} -- \eqref{eq:J_z_1_def}. Similar to the calculation of the radial and polar action in Section \ref{subsec:compactions}, we can use the fact that the actions are defined as integrals along the $\gamma_{r,z}$ loops in the phase space and write
\begin{align}
    \frac{\partial J_r^{(0)}}{\partial C_i} &= \frac{1}{2\pi} \oint \frac{\partial_{C_i} R_\text{so}(r)}{2\Delta\sqrt{R_\text{so}(r)}} \mathrm{d} r \,, \\
    \frac{\partial J_z^{(0)}}{\partial C_i} &= \frac{1}{2\pi} \oint \frac{\partial_{C_i} Z_\text{so}(z)}{2\Delta\sqrt{Z_\text{so}(z)}} \mathrm{d} z \,,
\end{align}
for $C_i = (E_\text{so},L_\text{so},K_\text{so})$ and similarly for the linear parts $J_{r,z}^{(1)}$. The $s_\parallel$ derivatives of $J_r$ and $J_z$ are simply $J_r^{(1)}/s_\parallel$ and $J_z^{(1)}/s_\parallel$. After the calculation of the derivatives, we obtain
\begin{align}
    \frac{\partial J_r^{(0)}}{\partial E_\text{so}} &= 
    \frac{1}{\pi} \int_{r_2}^{r_1} \frac{(r^2+a^2) ((r^2+a^2) E_\text{so} - a L_\text{so})}{\Delta \sqrt{R_\text{so}(r)}} \mathrm{d} r \,, \label{eq:dJr0dE}\\
    \frac{\partial J_r^{(0)}}{\partial L_\text{so}} &= - \frac{a}{\pi} \int_{r_2}^{r_1} \frac{(r^2+a^2) E_\text{so} - a L_\text{so}}{\Delta \sqrt{R_\text{so}(r)}} \mathrm{d} r \,, \label{eq:dJr0dL}\\
    \frac{\partial J_r^{(0)}}{\partial K_\text{so}} &= - \frac{1}{2 \pi} \int_{r_2}^{r_1} \frac{\mathrm{d} r}{\sqrt{R_\text{so}(r)}} \,, \label{eq:dJr0dK}\\
    \frac{\partial J_r^{(1)}}{\partial E_\text{so}} &= -\frac{s_\parallel \sqrt{K_\text{so}}}{\pi} \int_{r_2}^{r_1} \frac{\Delta (r^2+a^2)}{(R_\text{so}(r))^{3/2}} \mathrm{d} r \,, \label{eq:dJr1dE}\\
    \frac{\partial J_r^{(1)}}{\partial L_\text{so}} &= \frac{s_\parallel a\sqrt{K_\text{so}}}{\pi} \int_{r_2}^{r_1} \frac{\Delta}{(R_\text{so}(r))^{3/2}} \mathrm{d} r \,, \label{eq:dJr1dL}\\
    \frac{\partial J_r^{(1)}}{\partial K_\text{so}} &= -\frac{ s_\parallel }{2\pi\sqrt{K_\text{so}}} \int_{r_2}^{r_1} ((r^2+a^2) E_\text{so} - a L_\text{so}) \times \nonumber \\ &\phantom{=} \frac{(K_\text{so}-r^2)R_\text{so}(r) - K_\text{so}(K_\text{so}+r^2)\Delta }{(K_\text{so}+r^2)^2 (R_\text{so}(r))^{3/2}} \mathrm{d} r \,, \label{eq:dJr1dK}
\end{align}
for the radial action and 
\begin{align}
    \frac{\partial J_z^{(0)}}{\partial E_\text{so}} &= \frac{2a}{\pi} \int_{0}^{z_2} \frac{L_\text{so} - (1-z^2) a E_\text{so}}{\sqrt{Z_\text{so}(z)}} \mathrm{d} z \,, \label{eq:dJz0dE}\\
    \frac{\partial J_z^{(0)}}{\partial L_\text{so}} &= - \frac{2}{\pi} \int_{0}^{z_2} \frac{L_\text{so} - (1-z^2) a E_\text{so}}{(1-z^2) \sqrt{Z_\text{so}(z)}} \mathrm{d} z \,, \label{eq:dJz0dL}\\
    \frac{\partial J_z^{(0)}}{\partial K_\text{so}} &= \frac{1}{\pi} \int_{0}^{z_2} \frac{1}{\sqrt{Z_\text{so}(z)}} \mathrm{d} z \,, \label{eq:dJz0dK}\\
    \frac{\partial J_z^{(1)}}{\partial E_\text{so}} &= -\frac{2 s_\parallel a^2 \sqrt{K_\text{so}}}{\pi} \int_{0}^{z_2} \frac{(1-z^2)^2}{(Z_\text{so}(z))^{3/2}} \mathrm{d} z \,, \label{eq:dJz1dE}\\
    \frac{\partial J_z^{(1)}}{\partial L_\text{so}} &= \frac{2 s_\parallel a \sqrt{K_\text{so}}}{\pi} \int_{0}^{z_2} \frac{1-z^2}{(Z_\text{so}(z))^{3/2}} \mathrm{d} z \,, \label{eq:dJz1dL}\\  
    \frac{\partial J_z^{(1)}}{\partial K_\text{so}} &= - \frac{ s_\parallel a }{\pi\sqrt{K_\text{so}}} \int_{0}^{z_2} (L_\text{so} - (1-z^2) a E_\text{so}) \times \nonumber \\ &\phantom{=} \frac{(K_\text{so} + a^2 z^2) Z_\text{so}(z) + K_\text{so}(K_\text{so} - a^2 z^2)(1-z^2) }{(K_\text{so} - a^2 z^2)^2 (Z_\text{so}(z))^{3/2}} \mathrm{d} z \,, \label{eq:dJz1dK}
\end{align}
for the polar action.

Using this approach, we can find the frequencies $\Omega^r$, $\Omega^z$, and $\Omega^\phi$ up to the linear order in spin and the leading order of $\Omega^\psi$. The derivatives of the geodesic actions were expressed in Legendre form in \citet{Fujita:2009bp}, while the derivatives with respect to $s_\parallel$ can be found in Eqs.~\eqref{eq:J_r_1_eval} and \eqref{eq:J_z_1_eval}. In fact, the leading order of the precession frequency $\Omega^\psi$ agrees with the results of \citet{vandeMeent:2019cam} up to a different sign convention. 

The action integrals \eqref{eq:J_r_0_def}--\eqref{eq:J_z_1_def} and derivatives of the  zeroth-order (geodesic) actions \eqref{eq:dJr0dE}--\eqref{eq:dJr0dK} and \eqref{eq:dJz0dE}--\eqref{eq:dJz0dK} did not require Hadamard regularization. However, the derivatives of the spin corrections to the actions as given in equations \eqref{eq:dJr1dE}--\eqref{eq:dJr1dK} and \eqref{eq:dJz1dE}--\eqref{eq:dJz1dK} do require Hadamard's partie finie regularization because of the $(r - r_{1,2})^{-3/2}$ and $(z_2^2 - z^2)^{-3/2}$ divergences. Details about regularization and numerical verification can be found in Appendix \ref{app:freqs}. The final expression for the derivatives of $J_{r}^{(1)}$ and $J_{z}^{(1)}$ in Legendre form can be found in Eqs~\eqref{eq:Jr1_derivatives_results} and \eqref{eq:Jz1_derivatives_results}.

\section{Matching of Hamiltonians} \label{sec:match}

We will now compare the PN Hamiltonian for spinning binaries \cite{Tanay:2020gfb,Tanay:2021bff} with the expression for spinning test particles in Kerr space-time. 

The PN Hamiltonian generates motion parametrized by the time of inertial observers. For the spinning test particle, the corresponding Hamiltonian is $E_{\rm so}$ expressed as a function of actions on the mass-shell (four-velocity normalization) hypersurface, as discussed in Ref. \cite{Witzany:2018ahb}. As it does not seem possible to compute the inversion from $\vec{J}(\vec{C})$ to $\vec{C}(\vec{J})$ exactly, we will do so approximately in a PN expansion. To obtain the expanded function $E_{\rm so} (\vec{J})$, we first derive the PN expansion of actions as functions of the constants of motion, then iteratively invert for $E_{\rm so}$.

For the PN description of spinning binaries with finite mass ratios, the action Hamiltonian is also derived by inverting the relations between actions and energy, as discussed by \citet{Tanay:2021bff}. We correct a previous error, leading to a significant simplification of the Hamiltonian. We then take a spinning test particle limit of the PN Hamiltonian and compare the results obtained from the two approaches in order to obtain a dictionary between the actions of the two systems.

\subsection{PN limit of spinning test particle} \label{subsec:SpTPtoPN}

In the PN limit, the constants of motion of the spinning particle scale as $ E_{\rm so} \sim 1 + \mathcal{O}(\epsilon^2)$, $L_{\rm so} \sim \mathcal{O}(\epsilon^{-1})$, $K_{\rm so} \sim \mathcal{O}(\epsilon^{-2})$, and $s\sim s_\parallel \sim \mathcal{O}(1)$ where $\epsilon \sim v/c \sim \sqrt{Gm/(r c^2)}$ is the PN bookkeeping parameter. This leads to the parametrization by variables $\tilde{e},Y,\ell$ such that 
\begin{align}
    E_{\rm so} = 1 - \epsilon^2\frac{\tilde{e}^2}{2 \ell^2}\,,\; K_{\rm so} = \frac{1}{\epsilon^2} \ell^2, L_{\rm so} = \frac{1}{\epsilon} Y \ell\,, \label{eq:ellYe}
\end{align}
The variables $\tilde{e} ,Y$ are dimensionless, order 1 in the PN scaling and have the approximate meaning of a ``dual'' eccentricity (0 for highly eccentric, 1 for circular) and the sine of inclination, respectively. The only variable that comes with a PN scaling is $\ell$, it has dimensions of length (in $G=c=1$ units) and it has the approximate meaning of the specific angular momentum. 

Using the variables $\tilde{e},Y,\ell$, we expanded the roots $r_{1,2,3,4}$ and $z_{1,2}$ by iterating $R(r) = 0$ and $Z(z) =0$. The resulting expressions are in Appendix \ref{app:PNexp}. We then expanded formulas \eqref{eq:J_r_0_eval} and \eqref{eq:J_r_1_eval} to obtain
\begin{align}
    \label{eq:J_r_0_PNexp}
\begin{split}
  & J_r^{(0)} = 
    \left(\frac{1}{\tilde{e}} - 1\right) \ell 
    - \epsilon a Y \tilde{e}
    \\
    & \phantom{J_r^{(0)} =}
    + \epsilon^2 \frac{\left(M^2(24 - 15 \tilde{e}) + 4 a^2 (1+Y^2)\right)}{8 \ell}
    \\
    & \phantom{J_r^{(0)} =}
    - \epsilon^3 \frac{a Y\left(M^2(10 - \tilde{e}^2) + a^2(1+Y^2)\right)}{2 \ell^2} + \mathcal{O}(\epsilon^4) \,,
\end{split}
  \\
  \label{eq:J_r_1_PNexp}
\begin{split}
  & J_r^{(1)} = \epsilon s_\parallel 
  \left(
    1 - \epsilon \frac{M a Y}{\ell^2} - \epsilon^2 \frac{3 M^2 +a^2 (1+3 Y^2)}{2 \ell^2}
  \right)
  \\
  & \phantom{J_r^{(1)} =}
  + \mathcal{O}(\epsilon^4) \,.
\end{split}
\end{align}
Similarly, expanding \eqref{eq:J_z_0_eval} and \eqref{eq:J_z_1_eval} yields  
\begin{align}
\label{eq:J_z_0_PNexp}
\begin{split}
    & J_z^{(0)} = 
    \ell (1-|Y|) 
    +  \epsilon a Y
    - \epsilon^2 \frac{a^2 (1+Y^2)}{2 \ell}
    \\
    & \phantom{J_z^{(0)} =}
    +\epsilon^3 \frac{a Y\left(a^2(1+Y^2) -M^2 \tilde{e}^2\right)}{2 \ell^2}
    + \mathcal{O}(\epsilon^4)\,,
\end{split}
  \\
\label{eq:J_z_1_PNexp}
\begin{split}
  & J_z^{(1)} = \epsilon s_\parallel \left( \epsilon \frac{a Y}{\lambda} - \epsilon^2 \frac{a^2 (1+3 Y^2)}{2 \ell^2} \right) +\mathcal{O}(\epsilon^4)\,.
\end{split}
\end{align}
Formulas for the roots and actions up to $\mathcal{O}(\epsilon^6)$ are available in the supplemental Mathematica notebook.

We then proceeded to express $E_{\rm so} - 1$ (subtracting the rest-mass term as conventional) by substituting $\tilde{e} = \ell \sqrt{1-E_{\rm so}} $ 
and $Y = L_{\rm so}/\ell = J_\phi / \ell$, perturbatively eliminating $\ell$, and solving for $E_{\rm so} - 1$ to obtain (more details can be found in Appendix \ref{app:PNexp}) 
\begin{align}
\label{eq:HKs}
  & H_{\rm stp \to PN} \equiv E_{\rm so} - 1= H_{\rm Kg} + H_{\rm s}\,,
  \\
  \begin{split}
  & H_{\rm Kg} = 
    - \frac{M^2}{2 (J_r + J_z + |J_\phi|)^2} 
    \\
    & \phantom{H_{\rm Kg} = }
    - \epsilon^2 \frac{3 M^3 (8 J_r + 3(J_z + |J_\phi|))}{8(J_z + |J_\phi|)(J_r + J_z + |J_\phi|)^4}
    \\
    & \phantom{H_{\rm Kg} = }
    + \epsilon^3 \frac{2 M^4 a J_\phi}{(J_z + |J_\phi|)^3(J_r + J_z + |J_\phi|)^3}
    + \mathcal{O}(\epsilon^4)
    \,,
  \end{split}
  \\
  \begin{split}
  & H_{\rm s} = (J_\psi+s) M^2 \Big[
    \epsilon \frac{1}{(J_r + J_z + |J_\phi|)^3}
    \\
    &\phantom{H_{s} =}
    + \epsilon^3 \frac{3M^2\left[J_r^2 + 8J_r(J_z + |J_\phi|) +2 (J_z + |J_\phi|)^2\right]}{(J_z + |J_\phi|)^2(J_r + J_z + |J_\phi|)^5}
  \Big] 
  \\
  &\phantom{H_{s} =}
  + \mathcal{O}(\epsilon^4)\,.
  \end{split}
\end{align}
Here $H_{\rm Kg}$ is the Kerr geodesic part of the Hamiltonian and $H_{\rm s}$ is the secondary spin correction.

\subsection{Spinning-particle limit of finite mass-ratio Hamiltonian}
\label{subsec:PNtoSpTP}

%
As described by \citet{Tanay:2020gfb,Tanay:2021bff}, the Hamiltonian dynamics of spinning compact binaries with finite mass ratios is integrable at 1.5PN. These dynamics feature 8 degrees of freedom: 3 translational and 1 spin degree of freedom for each binary component. Three degrees of freedom can be combined into the total translational degrees of freedom of the center of mass. The dynamics in the center-of-mass frame are captured by the Hamiltonian $\mathcal{H}$ with 5 residual degrees of freedom and 5 associated actions $\mathcal{J}$ (denoted by calligraphic symbols to differentiate from spinning-particle ones). The 1.5PN order is the first order at which spins appear dynamically and is valid only to linear order in the spin vectors of the binary components $\vec{S}_1,\vec{S}_2$. Specifically, terms starting from
the ratio of $\sim S_1^2, S_2^2$ and $S_1 S_2$ with respect to orbital angular momentum squared are neglected. However, ratios of spin-proportional quantities \textit{between themselves} remain $\mathcal{O}(1)$ at this order.  

The first four actions are:
\begin{itemize}
    \item a radial action $\mathcal{J}_r$,
    \item the magnitude of orbital angular momentum $\vec{L} = \vec{p}\times \vec{x}$ of the binary $\mathcal{J}_L$,
    \item the magnitude of total angular momentum $\vec{L} + \vec{S}_1 + \vec{S}_2$ $\mathcal{J}_J$,
    \item and a single coordinate component of the total angular momentum $\mathcal{J}_z$. 
\end{itemize}
Since $\mathcal{J}_z$ depends on the coordinate axis choice and PN dynamics are rotation-invariant, it does not appear in the Hamiltonian. 

The first four actions are described in Ref. \cite{Tanay:2020gfb}. In Ref. \cite{Tanay:2021bff} the formula for the ``fifth'' action was derived, denoted as $\mathcal{J}_5$. In \citet{Tanay:2021bff} the formula for $\mathcal{J}_5$ was derived non-perturbatively, treating the 1.5PN Hamiltonian dynamics as an exact system. We revisited this work and found that when truncating $\mathcal{J}_5$ at leading PN order yields a simple piecewise continuous function\footnote{Note in particular that formula (47) for the PN-expanded version of $\mathcal{J}_5$ in Ref. \cite{Tanay:2021bff} contained a mistake that occluded this simplification.}
\begin{align}
    & \mathcal{J}_5 = 
        -\hat{L}\cdot \vec{S}_1 (1-\kappa_s) + \hat{L}\cdot \vec{S}_2 \kappa_s\,,
    \\
    & \kappa_s \equiv \Theta\left(|\hat{L}\times \vec{S}_1| - |\hat{L}\times \vec{S}_2|\right)\,,
\end{align}
where $\Theta$ is the Heaviside theta and we denote magnitudes of vectors as $|\vec{S}_{1,2}| = S_{1,2},\, |\vec{L}| = L$, and $\hat{L} = \vec{L}/L$ is the unit vector in the $\vec{L}$ direction. In other words, $\mathcal{J}_5$ only acquires values either $-\vec{L}\cdot \vec{S}_1/L$ or $\vec{L}\cdot \vec{S}_2/L$. Which of these values is picked is decided by finding the part of the spins orthogonal to orbital angular momentum, $\hat{L} \times \vec{S}_{1,2}$, and checking which one has a greater magnitude. Note in particular that $\mathcal{J}_5$ depends only on projections of spin into the $\vec{L}$ direction, not on the $L$ magnitude.

The discontinuities in the actions as functions of phase space coordinates are not strictly necessary. One can transform from $\mathcal{J}_J, \mathcal{J}_5$ to an alternative action pair as follows:
\begin{align}
    & \mathcal{J}_{S1} = (\mathcal{J}_J - \mathcal{J}_L)\kappa_s -  \mathcal{J}_5\,,
    \\
    &\mathcal{J}_{S2} = (\mathcal{J}_J - \mathcal{J}_L)(1-\kappa_s) + \mathcal{J}_5\,.
\end{align}
This action set satisfies the simpler continuous relations $\mathcal{J}_{S1} = \vec{L}\cdot \vec{S}_1/L$, $\mathcal{J}_{S2} = \vec{L}\cdot \vec{S}_2/L$ at leading order when $\vec{L} \gg \vec{S}_{1,2}$. However, the actions $\mathcal{J}_J, \mathcal{J}_5$ have the advantage of having a definition that is symmetric and antisymmetric with respect to $\vec{S}_1 \leftrightarrow \vec{S}_2$ switching. Additionally, $\mathcal{J}_5$ vanishes when either spin is zero, causing the Hamiltonian to automatically become a manifestly 3-DOF Hamiltonian when relevant. Thus, the $\mathcal{J}_J$ and $\mathcal{J}_5$ actions are better suited for an ``Effective one body'' type of approach.

To invert the Hamiltonian, we utilize the key formula (38) for the radial action from Ref. \cite{Tanay:2020gfb}
\begin{align}
\begin{split}
    & \mathcal{J}_r = 
        -\mathcal{J}_L 
		+ \frac{M_{\rm t} \mu^{3/2}}{\sqrt{-2\mathcal{H}}} 
        \\
        & \phantom{\mathcal{J}_r =}
		+ \epsilon^2 M_{\rm t} \left[ 
			\frac{3 M_{\rm t} \mu^2}{\mathcal{J}_L} 
			+ \frac{\sqrt{-\mathcal{H}} \mu^{1/2}(\nu - 15)}{\sqrt{32}} 
		\right]
        \\
        & \phantom{\mathcal{J}_r =}
        - \epsilon^3 \frac{2 \mu^3}{\mathcal{J}_L^3} \vec{L}\cdot \vec{S}_{\rm eff}
		+ \mathcal{O}(\epsilon^4)\,,
\end{split}
	\\
	& \vec{S}_{\rm eff} \equiv \sigma_1 \vec{S}_1 + \sigma_2 \vec{S}_2\,,
	\\
	& \sigma_1 \equiv 1 + \frac{3 m_2}{4 m_1}\,,\; \sigma_2 \equiv 1 + \frac{3 m_1}{4 m_2} \,,
\end{align}
where $m_{1,2}$ are the component masses, $M_{\rm t} = m_1 + m_2$ is the total mass of the binary, and $\mu = m_1m_2/M_{\rm t}$ is the reduced mass of the binary.

Using the previously defined actions, we obviously have 
\begin{align}
\begin{split}
    \vec{L}\cdot \vec{S}_{\rm eff} 
    & = \mathcal{J}_L(\sigma_1 \mathcal{J}_{S1} +\sigma_2 \mathcal{J}_{S2}) \\
    & = \mathcal{J}_L \big[
            \left(\sigma_1\kappa_s + \sigma_2 (1-\kappa_s) \right) (\mathcal{J}_J - \mathcal{J}_L)
            \\
            & \phantom{= \mathcal{J}_L \big[}\,
            + (\sigma_2 - \sigma_1) \mathcal{J}_5
        \big] \, .
\end{split}
\end{align}  
Inverting the formula for $\mathcal{J}_r$ gives the Hamiltonian as
 \begin{align} \label{eq:HLSeff}
 \begin{split}
    &\mathcal{H} =
	-\frac{M_{\rm t}^2 \mu^3}{2 (\mathcal{J}_r + \mathcal{J}_L)^2}
    \\
    & 
	+ \epsilon^2 \frac{M_{\rm t}^4 \mu^5}{8}\left[
		\frac{15-\nu}{(\mathcal{J}_r + \mathcal{J}_L)^4}
		- \frac{24}{\mathcal{J}_L(\mathcal{J}_r + \mathcal{J}_L)^3}
	\right]
    \\
    & 
	+ \epsilon^3 \frac{2 M_{\rm t}^3 \mu^6}{\mathcal{J}_L^2(\mathcal{J}_r + \mathcal{J}_L)^3} \big[
            \left(\sigma_1\kappa_s + \sigma_2 (1-\kappa_s) \right) (\mathcal{J}_J - \mathcal{J}_L)
            \\
            & \phantom{+ \epsilon^3 \mathcal{J}_L^2(\mathcal{J}_r + \mathcal{J}_L)^3 \big[}\;\,
            + (\sigma_2 - \sigma_1) \mathcal{J}_5
        \big] + \mathcal{O}(\epsilon^4)\,.
 \end{split}
 \end{align}
This formula is the first time the closed-form action-angle Hamiltonian for eccentric and precessing spinning binaries at 1.5PN appears in the literature. Note that the result does not depend on the spin magnitudes, only on the actions $\mathcal{J}_{J} \in [0,L+S_{1}+ S_{2}]$ and $\mathcal{J}_5 \in [- \text{max}(S_1,S_2),\text{max}(S_1,S_2)]$.

So far, we have made no assumptions about the mass ratio. We now consider the ``spinning test particle limit'', expanding to zeroth order in $m_2\to 0$ and first order in specific secondary spin $\vec{s}_2 = \vec{S}_2/m_2$. This limit eliminates gravitational self-force terms, which are absent in the previously obtained Hamiltonian of the spinning test particle. 

However, before doing so, one must realize that in the test particle limit $S_1 \gg L$, so the limit \textit{does not} commute with the PN limit. For example, the expression 
\begin{align}
    \mathcal{J}_L\left[(\mathcal{J}_J - \mathcal{J}_L)\kappa_s -  \mathcal{J}_5\right] \to \vec{L} \cdot \vec{S}_1 \text{ when } L\gg S_1\,, 
\end{align}
but this is not true in the $S_1 \gg L$ limit. To amend this, we add subleading PN terms to obtain an expression for $\vec{L}\cdot \vec{S}_1$ valid in both limits:
\begin{align}
\begin{split}
    & (\mathcal{J}_L+S_1)\left[(\mathcal{J}_J - \mathcal{J}_L)\kappa_s -  \mathcal{J}_5\right] - S_1^2 \to \vec{L} \cdot \vec{S}_1 
    \\
    & \text{ when } L\gg S_1 \text{ or } S_1\gg L\gg S_2\,.
\end{split}
\end{align}
Replacing $\vec{L}\cdot\vec{S}_1$ with this regularized expression when formulating the Hamiltonian yields a formal 2PN regularization of the Hamiltonian
\begin{align}
\begin{split}
    &\mathcal{H}_{\rm reg} = 
     \mathcal{H} 
    \\
     &\quad\; + \epsilon^4 \frac{2 M_{\rm t}^3 \mu^6}{\mathcal{J}_L^2(\mathcal{J}_r + \mathcal{J}_L)^3} 
     \left[
    S_1 \sigma_1\kappa_s (\mathcal{J}_J - \mathcal{J}_L) - S_1^2\right].
\end{split}
\end{align}

Other regularization options are possible, in particular those symmetrized in $S_1 \leftrightarrow S_2$. However, they always seem to yield the same results in the spinning test particle limit, provided the $\vec{L}\cdot \vec{S}_1$ expression is recovered in both $L\gg S_1$ and $S_1 \gg L$ limits. 

To facilitate the limit and subsequent comparison, we rescale to mass-specific variables
\begin{align}
    I_{r,L,J,5} \equiv \frac{\mathcal{J}_{r,L,J,5}}{\mu} \,,\; H_{\rm PN} \equiv \frac{\mathcal{H}_{\rm reg}}{\mu} \,.
\end{align}
For instance, $I_L$ denotes the magnitude of the specific angular-momentum vector $\vec{l} \equiv \vec{L}/\mu$.
Consequently, we have
\begin{align}
    \frac{\partial H_{\rm PN}}{\partial I_{r,L,J,5}} = \frac{\partial \mathcal{H}}{\partial \mathcal{J}_{r,L,J,5}}\,.
\end{align}
In other words, the specific Hamiltonian $H_{\rm PN}(I)$ where $I$ are understood as action variables canonically conjugate to the original angles generates the same motion as $\mathcal{H}(\mathcal{J})$. 

We can also change the origin of the actions. For the variable $I_{J}$ its magnitude is dominated by a divergent term $S_1/\mu$ when $S_1\gg L$. To obtain a regular action in the $m_2 \to 0$ limit, we define
\begin{align}
    I_{\Delta J} 
    & =  I_{J} - \frac{S_1}{\mu} 
    \\
    & = \frac{1}{\mu}\left(\sqrt{L^2 + S_1^2 + S_2^2 + 2 \vec{L}\cdot\vec{S}_1 + 2 \vec{L}\cdot\vec{S}_2 + 2 \vec{S}_1 \cdot \vec{S}_2} -S_1\right)
    \\
    & = \vec{l}\cdot \hat{S}_1 + \vec{s}_2 \cdot \hat{S}_1 + \mathcal{O}(m_2)\,, \label{eq:IDeltaJtp}
\end{align}
where we have switched to mass-specific vectors $\vec{l} = \vec{L}/\mu, \vec{s}_2 = \vec{S}_2/\mu$ and $\hat{S}_1 = \vec{S}_1/S_1$.

When using the relations above and when discarding all of the $\mathcal{O}(m_2)$ terms while keeping $\mathcal{O}(s_2)$ terms, we obtain the Hamiltonian

\begin{align}
\label{eq:HPNstp}
\begin{split}
    & H_{\rm PN \to stp} = 
    -\frac{m_1^2}{(I_r + I_L)^2} 
    - \epsilon^2 \frac{ 3m_1^4(8 I_r + 3 I_L)}{8 I_L (I_r + I_L)^4}
    \\
    & \phantom{H_{\rm PN \to stp} = }
    + \epsilon^3\frac{2 m_1^4 S_1 I_{\Delta J}}{I_L^3 (I_r + I_L)^3} 
    + \epsilon^3 \frac{3 m_1^4 I_{5}}{2I_L^2 (I_r + I_L)^3}
    \\
    & \phantom{H_{\rm PN \to stp} = }
    + \mathcal{O}(\epsilon^4,m_2)
    \,,
\end{split}
\end{align}
where we ignored the edge case $|\vec{L}\times\vec{S}_2| > |\vec{L}\times \vec{S}_1|$ in the $S_1\gg L \gg S_2$ limit. 

\subsection{Comparison of Hamiltonians} \label{subsec:finalmatch}
Let us examine the limiting expressions \eqref{eq:HPNstp} and \eqref{eq:HKs} obtained in the overlapping region of validity of the PN and spinning test particle limit. 

First consider the expression for the nonspinning secondary, or $s_2=s_\parallel=s=J_\psi=I_{5}=0$. From the 0 and 1PN orders, we find
$I_r = J_r + \mathcal{O}(s_2)$,
$I_{L} = J_z + |J_\phi| + \mathcal{O}(s_2)$ and $M = m_1 + \mathcal{O}(m_2,s_2)$. From the 1.5PN term, we obtain $I_{\Delta J} = J_\phi + \mathcal{O}(s_2)$ and $M a = S_1 + \mathcal{O}(m_2,s_2)$. The $I_{\Delta J}$-$J_\phi$ match has a natural interpretation, as $I_{\Delta J}$ is canonically conjugate to rotations of the entire binary around the total angular momentum axis \cite{Tanay:2020gfb}, and in the test particle limit such rotations reduce to $\phi$ rotations around the spin axis of the heavy Kerr black hole.   

Now, let us examine the secondary spin corrections. Since the only $\mathcal{O}(s_2)$ action is $J_\psi$ and the actions have to be related by integer addition, we substitute the Ansatz
\begin{align}
    & I_r = J_r + n_r (J_\psi + s)\,,\\
    & I_{L} = J_z + |J_\phi| + n_L (J_\psi+ s) \,,\\
    & I_{\Delta J} = J_\phi + n_{\Delta J} (J_\psi+s) \,,\\
    & I_{5} = n_s (J_\psi + s)\,,
\end{align}
where $n_{r,L,\Delta J,s}$ are integers. We used the particular combination $s_\parallel =J_\psi + s$, as it is the only spin-dependent quantity that appears in observables (see Refs. \cite{Witzany:2019nml,Skoupy:2023lih,Piovano:2024yks}). Requiring that $H_{\rm PN \to stp} = H_{\rm stp \to PN}$ order by order in $\epsilon$ leads to the matching conditions
\begin{align}
    n_r + n_L = -1\,, \quad (1+2n_L) n_s = -1\,.
\end{align}
Furthermore, the requirement that the absolute value of the determinant of the transform is 1 leads to the condition
\begin{align}
    |n_s| = 1 \,.
\end{align}
Interestingly, $n_{\Delta J}$ does not appear in the conditions above, so we obtain no constraint on it at this level. 

 The action $J_\psi = s_{\parallel} -s$, where $s_\parallel$ is defined as $s^\mu l_\mu/\sqrt{l^\nu l_\nu}$ through the spin vector of the particle $s^\mu$ and a specific orbital angular-momentum vector $l^\mu = Y^{\mu\nu} u_\nu$, where $u^\mu$ is the four-velocity and $Y^{\mu\nu}$ the Killing-Yano tensor of the Kerr space-time (see ref. \cite{Witzany:2019nml}). We then see that $I_5$ and $s_\parallel =J_\psi+s$ obviously have the same meaning in the overlapping limits. We thus conclude that $n_s = 1$. This leads to $n_r = 0$ and $n_L = -1$.

The case of $n_{\Delta J}$ is more tricky. The $\phi$ axis in Kerr space-time is defined exclusively by the primary spin, and this is also true for the definition of $J_\phi$. Its meaning can therefore be conjectured to be $J_\phi \sim (\vec{L}+ \vec{S}_2) \cdot \vec{S}_1/(|\vec{S}_1|m_2)$. We get the exact same meaning for $I_{\Delta J}$ when considering the limit $S_1 \gg \vec{L}, S_2$ at leading order as can be seen in equation \eqref{eq:IDeltaJtp}. Therefore, we conjecture $n_{\Delta J} = 0$. 

There is another geometrical argument to show that $n_{\Delta J} = 0$. As discussed by \citet{Tanay:2020gfb}, the Poisson bracket with the magnitude of the total angular momentum $\{\; ,|\vec{L} + \vec{S}_1 + \vec{S}_2|\}$ generates a rotation of the entire binary about the total angular momentum axis, including its spin vectors. This is also true for $I_{\Delta J}$ since $|\vec{S}_1|$ and $\mu$ commute with all phase-space variables. In the test-particle limit of the PN description, this transformation reduces to the rotation of the orbit and the spin vector $\vec{S}_2$ about $\vec{S}_1$. Is this reproduced by the relativistic computation? The action $J_\phi$ can be expressed in the original coordinates $\pi_\mu, s^{AB}$ as the constant of motion generated by the azimuthal-angle Killing vector $\xi^{(\phi)}_\mu$ \cite{Dixon:1970I,Witzany:2018ahb}
\begin{align}
\begin{split}
    J_\phi 
    & = \xi^{(\phi) \mu} p_\mu + \frac{1}{2} \xi^{(\phi)}_{\kappa;\lambda} s^{\kappa\lambda}
    \\
    & = \xi^{(\phi) \mu} \pi_\mu + \frac{1}{2} \left(\xi^{(\phi)}_{\kappa;\lambda} e^\kappa_A e^\lambda_B - \xi^{(\phi) \mu} e^\gamma_{A;\mu}e_{B\gamma} \right) s^{AB}\,.
\end{split}   
\end{align}
For $\phi$-independent tetrads $e^A_\mu$ such as the Marck congruence tetrad, the second term in the second row vanishes, and one obtains simply $J_\phi = \xi^{(\phi) \mu} \pi_\mu = \pi_\phi$ \cite{Witzany:2018ahb}. However, for other tetrads, such as Cartesian-like tetrads, the second term becomes nonzero. We have verified that in such a case the Poisson bracket $\{\; ,J_\phi\}$ generates a rotation of both the orbital configuration and the Cartesian-like components of the spin tensor around the spin axis of the massive black hole. As such, $J_\phi$ and $I_{\Delta J}$ correspond to the same loop in phase space even when the secondary spin is nonzero.

In summary, the following correspondence can be established between the PN and spinning-particle dynamics
\begin{align}
    & I_r = J_r \,, \label{eq:IrJr}\\
    & I_{L} = J_z + |J_\phi|- (J_\psi +s) \,, \label{eq:ILlat}\\
    & I_{\Delta J} = J_\phi  \,,\\
    & I_{5} = J_\psi +s  \,. \label{eq:IsJps}
\end{align}
 As discussed, these relations should apply to dynamics at finite mass ratios and higher PN orders, as they cannot be continuously deformed. 

\subsection{Implications of the matching} \label{subsec:matchimplications}

From the matching in equations \eqref{eq:IrJr}--\eqref{eq:IsJps}, we can automatically derive the relations for angle variables. This is because if one set of actions is related to another by an integer matrix $\mathbb{M}$ as $\vec{I} = \mathbb{M} \cdot \vec{J}$, then the angles $\vec{\xi}$ conjugate to $\vec{I}$ fulfill a complementary relation $\vec{\xi} = (\mathbb{M}^{T})^{-1} \cdot \vec{\eta}$, where $\vec{\eta}$  are conjugate to $\vec{J}$ \cite{fasano2006analytical}. In our case this leads to

\begin{align}
    & \xi^r = \eta^r \,, \label{eq:xir}\\
    & \xi^{L} = \eta^z \,, \label{eq:xiL}\\
    & \xi^{\Delta J} = \eta^\phi - {\rm sign}(J_\phi)\eta^z  \label{eq:xiDeltaJ}\,,\\
    & \xi^{5} =  \eta^\psi + \eta^z\,. \label{eq:xiS}
\end{align}
 The equalities indicate relations expected to hold in any integrable system that smoothly connects the spinning test particle limit to the dynamics of spinning binaries at finite mass ratios. 

The angles are complicated phase-space functions, making these relations challenging to apply in practice. Nonetheless, the same relations hold for the fundamental frequencies of the motion $\vec{\Omega} = \mathrm{d} \vec{\eta}/ \mathrm{d} t$ and $\vec{\tilde{\Omega}} = \mathrm{d} \vec{\xi}/ \mathrm{d} t$:
\begin{align}
    & \tilde{\Omega}^r = \Omega^r \,, \label{eq:Omr}\\
    & \tilde{\Omega}^{L} = \Omega^z \,, \label{eq:OmL}\\
    & \tilde{\Omega}^{\Delta J} = \Omega^\phi -  {\rm sign}(J_\phi) \Omega^z  \label{eq:OmDeltaJ}\,,\\
    & \tilde{\Omega}^{5} =  \Omega^\psi + \Omega^z\,. \label{eq:OmS}
\end{align}
Again, these equalities should hold for dynamics smoothly interpolating the two limiting cases. Their advantage is that fundamental frequencies of motion can be extracted in a much more coordinate independent fashion by Fourier analysis of the dynamical system. 

For example, in a numerical relativity simulation, $\tilde{\Omega}^r$ represents the dominant frequency in the Fourier decomposition of any reasonable radial-distance coordinate.\footnote{By ``reasonable'', we mean here the requirement that the definition does not exhibit any singular behavior and reproduces the correct relations to some order of PN dynamics (including at least leading order or 0PN).} $\tilde{\Omega}^L$ is then the dominant frequency for angular motion in the orbital plane (as observed in a suitable coprecessing or ``twisted'' frame \cite{Schmidt:2010it,Schmidt:2012rh}). The frequency $\tilde{\Omega}^{\Delta J}$ indicates the dominant frequency driving a reasonably defined orbital angular-momentum vector. Alternatively, $\tilde{\Omega}^{\Delta J}$ can be characterized as the frequency of the precession of the nodes. The frequency $\tilde{\Omega}^5$ has a less clear meaning and would likely require matching the PN evolution of Ref. \cite{Samanta:2022yfe} to numerical relativity to understand its extraction procedure.

Before concluding this section, we note that lattice transformations between actions permit \textit{discontinuities} that may not be reflected in equations \eqref{eq:IrJr}-\eqref{eq:IsJps}.  For example, the lattice transform leading to $I_L$ in eq. \eqref{eq:ILlat} has a jump at $J_\phi = 0$ (reflected by the absolute value in the $|J_\phi|$ term) due to a branch cut that is known to appear already when defining similar relations at the Newtonian order. Another example is the Schwarzschild limit of Kerr space-time. At finite $a$, the axis defining $J_\phi$ is tied to the total angular momentum of the system (dominated by the spin of the heavy central black hole) and $J_\phi$ is thus related to $\mathcal{J}_J$. At exactly $a=0$, the axis that defines $J_\phi$ loses physical meaning and instead we should identify $J_\phi$ with $\mathcal{J}_z$. Additionally, $\mathcal{J}_5$ as a function of the original binary parameters and phase-space coordinates exhibits a discontinuity along $|\vec{L}\times \vec{S}_1| = |\vec{L}\times \vec{S}_2|$. It is possible that when using our results outside of the spinning test particle limit, the lattice transformation in equations \eqref{eq:IrJr} to \eqref{eq:IsJps} will need to be refined by discrete jumps due to these considerations. 

\section{Discussion and Outlooks} \label{sec:discussion}

 This work presents a novel gauge-invariant match between the degrees of freedom of spinning compact binaries and spinning test particles in Kerr space-time. This generalizes the matching by \citet{Buonanno:1998gg} for non-spinning binaries and offers a geometrical counterpart to the pragmatic matchings for spinning binaries by \citet{Damour:2001tu} and \citet{Barausse:2009xi}. Currently, it serves as a proof of principle for matching in more general cases to follow this work. The next step will be to construct action variables for spinning particles in an effective deformed Kerr metric. Similar considerations as in Refs. \cite{Buonanno:1998gg,Damour:2001tu,Barausse:2009xi} should then enable the development of geometrically motivated effective one-body models for fully precessing spinning binary inspirals.

A key issue during the matching procedure was the non-commutation of the test-particle and PN limits. This occurred because orbital angular momentum $\vec{L}$ dominated over primary spin $\vec{S}_1$ in the PN limit but was sub-dominant in the test-particle limit. In Section \ref{subsec:PNtoSpTP} we addressed this by employing a regularized expression for $\vec{L}\cdot \vec{S}_1$ valid in both the $L\gg S_1$ and $S_1 \gg L$ limits. The resulting regularization terms were formally of 2PN order. Could we devise a systematic method for taking the double limit by extending the action variables to 2PN order, where physical spin-squared effects appear? This is also a goal for future work. 

Going to 2PN may indicate that the aforementioned regularization is not allowed at that order, and that the test particle actions cannot be unambiguously matched to finite mass ratio actions beyond 1.5PN. Instead, it is possible that one will need to obtain finite-mass-ratio corrections to the test particle side of the correspondence to resolve this issue. This is because in the test particle limit the frame is always strictly fixed to the static primary spin $\vec{S}_1 \sim M a \hat{z}$ and there is no known set of phase-space functions that would represent the generators of the fundamental 3D rotation symmetries $\vec{J}$. In other words, a part of the dynamics is ``washed out'' by the test particle limit. However, the metric perturbation sourced by the test particle also cause perturbations to asymptotic inertial frames, which will generally misalign them with respect to primary spin. In these frames, the primary spin appears dynamical.  Additionally, \citet{Blanco:2022mgd,Blanco:2023jxf} showed that from linear order in the mass ratio, the symplectic structure of the Hamiltonian description of the self-forced particle must be deformed away from that of test particles. A deeper understanding of both inertial frame perturbations and the geometry of the deformed phase space should help recover the ``washed out" dynamics of the primary spin.

One of the main results of this paper is the expression for the actions of spinning test particles in Kerr space-time and the derived fundamental frequencies of motion. Notably, while \citet{Schmidt:2002qk} provided the integral expression for Kerr geodesics, our formulas $J_r^{(0)},J_z^{(0)}$ in equations \eqref{eq:J_r_0_eval} and \eqref{eq:J_z_0_eval} represent the first closed-form expression of Kerr geodesic actions in terms of Legendre elliptic integrals. The fundamental frequencies of motion are a crucial ingredient for computing the gravitational perturbations sourced by the particle in the Fourier domain \cite{Drasco:2003ky}. In Ref. \cite{Skoupy:2023lih} gravitational-wave fluxes sourced by spinning particles were computed while the frequencies obtained by the numerical method of \citet{Drummond:2022efc,Drummond:2022xej} were used. More recently, \citet{Piovano:2024yks} used an alternative method for the computation while using a semi-analytical evaluation of the frequencies through the Hamilton-Jacobi formalism of Ref. \cite{Witzany:2019nml}. We verified that our fully closed-form expressions for the frequencies derived here agree with the results of these previous numerical and/or semi-analytical computations. As such, they can be readily implemented in the aforementioned flux computations to make them more efficient and accurate. Another possible application for the future is to use the derivatives of the actions to compute the ``flux'' of the R{\"udiger}-Carter constant from Teukolsky amplitudes, as proposed by \citet{Grant:2024ivt}.

While working on this paper, we learned of the independent work of \citet{Gonzo:2024zxo} with results that seem closely related to ours. The authors took the solution of the Hamilton-Jacobi equation for spinning test particles in Kerr space-time from Ref. \cite{Witzany:2019nml} and dropped the Ricci-rotation terms. Nevertheless, the Ricci-rotation terms are necessary to produce correct orbital shapes in Boyer-Lindquist coordinates, as verified directly by \citet{Piovano:2024yks}, so the action of \citet{Gonzo:2024zxo} will produce incorrect results for local trajectories. This being said, various invariant or global observables obtained from this action may still be correct due to the Hadamard regularization argument of \citet{Damour:1988mr} discussed in our Section \ref{subsec:compactions}. Gonzo \& Shi then computed actions for scattering orbits and used the boundary-to-bound map of \citet{Kalin:2019rwq,Kalin:2019inp} to obtain a proposal for the actions and fundamental frequencies of bound motion of spinning test particles expressed using Lauricella hypergeometric functions. We reexpressed their formulas for the polar action in Legendre form and obtained results equivalent to our eqns. \eqref{eq:J_z_0_eval} and \eqref{eq:J_z_1_eval}. However, the radial action of Ref. \cite{Gonzo:2024zxo} is defined by a complex continuation of a difference of two infinite actions corresponding to scattering, which was difficult to re-express in a manifestly finite and closed form equivalent to ours.  We leave a more detailed analysis of the relation of our results with those of \citet{Gonzo:2024zxo} for future work.

\begin{acknowledgments}
%
%
We would like to thank Gabriel Piovano and Riccardo Gonzo for useful feedback on the paper.
VW and VS thank the Charles University \textit{Primus} Research Program 23/SCI/017 for support. The work of LCS was supported by NSF CAREER Award PHY--2047382 and a Sloan Foundation Research Fellowship.
ST was supported by the PSL Postdoctoral Fellowship.
\end{acknowledgments}

\appendix

\section{Reduction of actions into Legendre form} \label{app:leg}

In this section, we will drop the ``so'' subscripts on $R_{\rm so}, Z_{\rm so}$ and understand it as implicit. Furthermore, we will use the fact that $C_{\rm c} = C_{\rm so} + \mathcal{O}(s,s_\parallel)$ for $C=E,L,K$, so we can use the ``so'' constants everywhere in the spin corrections $J_y^{(1)}$.

\subsection{Radial geodesic action}
Starting with the $J_{r}^{(0)}$ integral in eq. \eqref{eq:J_r_0_def}, we put the integrand into standard
Legendre forms as follows.  Transform to a new variable $\rho$,
\begin{align}
  \label{eq:rho_def}
  \rho^{2} &\equiv \frac{(r-r_{2})(r_{1}-r_{3})}{(r_{1}-r_{2})(r-r_{3})}
  \,,\; 
  r = \frac{r_{3}(r_{1}-r_{2})\rho^{2}-r_{2}(r_{1}-r_{3})}{(r_{1}-r_{2})\rho^{2}-(r_{1}-r_{3})}
  \,.
\end{align}
Note that when integrating the radial equation in closed
form~\cite{Fujita:2009bp,vandeMeent:2019cam}, this $\rho$ coordinate
will be a Jacobi elliptic sine function $\mathrm{sn}()$, which we will encounter below as well.  Note that when $r\in[r_{2},r_{1}]$, we have
$\rho\in[0,1]$, with equality at the endpoints as ordered.  We need
the following Jacobian of this transformation,
\begin{align}
  \label{eq:dr_drho}
  \frac{dr}{d\rho} = \frac{2 (r_{1} - r_{2}) (r_{1} - r_{3}) (r_{2} - r_{3}) \rho}{((r_{1}-r_{2}) \rho^2 - (r_{1}-r_{3}))^2}
  \,.
\end{align}
In terms of this new variable, our radical becomes
\begin{align}
\begin{split}
  \sqrt{{R}} = 
    & (r_{1} - r_{2}) (r_{2} - r_{3}) (r_{1} - r_{3})^{3/2} (r_{2} - r_{4})^{1/2} \times
    \\
    & \sqrt{1 - E_{\rm so}^2}
    \frac{\rho\sqrt{(1 - \rho^2) (1 -
      k_{r}^{2} \rho^2)}}{((r_{1}-r_{2}) \rho^2 - (r_{1}-r_{3}))^2}
    \,,
\end{split}
\end{align}
where we have defined the elliptic modulus
\begin{align}
  k_{r}^{2} \equiv \frac{(r_{1} - r_{2}) (r_{3} - r_{4})}{(r_{1} - r_{3}) (r_{2} - r_{4})}
  \,.
\end{align}
Notice that $0 \le k_{r}^{2}$ when $r_1>r_2>r_3$ are positive, which is true for generic bound motion and $k_r = 0$ for circular orbits. In the PN expansion one can show that $k_r \lesssim v^2/c^2 \ll 1$ (see Appendix \ref{app:PNexp}). The elliptic integral expressions have logarithmic singularities for $k_r^2 = 1$, which occurs at the separatrix of bound orbits when $r_2 \to r_3$. The $r_2 \to r_3$ bound orbits are also known as the marginally stable orbits since small perturbations send them into plunging motion into the black hole \cite{Stein:2019buj}. Hence, we will focus on the case $0\leq k_r^2 <1$ of stable bound motion here; other cases could be analyzed by methods similar to those of Refs \cite{Stein:2019buj,Compere:2021bkk}, but are out of the scope of our paper. 
Note also that van de Meent~\cite{vandeMeent:2019cam} defined his $k_{r}$
as our $k_{r}^{2}$ (his should rather have been called the elliptic
\emph{parameter} $m_{r}$, rather than the modulus).  Now, to
transform $\Delta$, we define the shorthands
\begin{align}
  \label{eq:rho_plus_minus_def}
  \rho_{\pm}^{2} \equiv \frac{(r_{\pm}-r_{2})(r_{1}-r_{3})}{(r_{1}-r_{2})(r_{\pm}-r_{3})}
  \,,
\end{align}
which are simply the value that $\rho$ takes when evaluated at
$r=r_{\pm}$.  Now $\Delta$ is expressed in the $\rho$ coordinate as
\begin{align}
  \Delta = (r_{1} - r_{2})^2 (r_{3} - r_{-}) (r_{3} - r_{+})
  \frac
  {(\rho^2 - \rho_{+}^2) (\rho^2 - \rho_{-}^2)}
  {((r_{1} - r_{2}) \rho^2 - (r_{1} - r_{3}))^2}
  \,.
\end{align}
Finally, assembling $J_{r}^{(0)}$ in the $\rho$ coordinate, we get
\begin{align}
\begin{split}
  J_{r}^{(0)} 
    =& \frac{1}{\pi} \int_{0}^{1} \frac{\sqrt{R(\rho)}}{\Delta(\rho)} \frac{dr}{d\rho} d\rho \,,\\
    =& \frac{2}{\pi} 
    \frac
        {\sqrt{1 - E_{\rm so}^2} \sqrt{(r_{1} - r_{3}) (r_{2} - r_{4})}(r_{2} - r_{3})^{2} }
        {(r_{3} - r_{-}) (r_{3} - r_{+}) } \times
    \\
    &
    \int_{0}^{1}
    \frac
        {\sqrt{(1 - \rho^2) (1 - k_{r}^{2} \rho^2)}}
        {(\rho^2 - \rho_{+}^2) (\rho^2 - \rho_{-}^2)}
    \frac{\rho^{2}}{(\alpha_{r}^{2} \rho^2 - 1)^2}
    d\rho\,,
\end{split}
\end{align}
where we have also defined
\begin{align}
  \alpha_{r}^{2} = \frac{r_{1}-r_{2}}{r_{1}-r_{3}}
  \,.
\end{align}
In the Legendre canonical form, the radical is always in the
denominator,
\begin{align}
\begin{split}
  & \frac
  {\sqrt{(1 - \rho^2) (1 - k_{r}^{2} \rho^2)}}
  {(\rho^2 - \rho_{+}^2) (\rho^2 - \rho_{-}^2)}
  \frac{\rho^{2}}{(\alpha_{r}^{2} \rho^2 - 1)^2}
  \\ 
  & =
  \frac{1}{\sqrt{(1 - \rho^2) (1 - k_{r}^{2} \rho^2)}}
  \frac
  {(1 - \rho^2) (1 - k_{r}^{2} \rho^2)\rho^{2}}
  {(\rho^2 - \rho_{+}^2) (\rho^2 - \rho_{-}^2)(\alpha_{r}^{2} \rho^2 - 1)^2}
  \,.
\end{split}
\end{align}
Next we do a partial fraction decomposition of the rational factor.
However, noticing that only even powers of $\rho$ appear, we do not
perform a complete partial fractions decomposition.  Instead, we
decompose as
\begin{align}
\begin{split}
  &\frac
  {(1 - \rho^2) (1 - k_{r}^{2} \rho^2)\rho^{2}}
  {(\rho^2 - \rho_{+}^2) (\rho^2 - \rho_{-}^2)(\alpha_{r}^{2} \rho^2 - 1)^2}
  \\& =
  \frac{\gamma_{r,0,+}}{\rho^{2}-\rho_{+}^{2}}
  + \frac{\gamma_{r,0,-}}{\rho^{2}-\rho_{-}^{2}}
  + \frac{\gamma_{r,0,\alpha}}{\alpha_{r}^{2}\rho^{2}-1}
  + \frac{\delta_{r,0,\alpha}}{(\alpha_{r}^{2}\rho^{2}-1)^{2}}
  \,,
\end{split}
\end{align}
where we have defined the shorthands
\begin{align}
  \label{eq:gamma_plusminus}
  \gamma_{r,0,\pm} &\equiv
  \mp
  \frac
  {(r_{1} - r_{\pm}) (r_{2} - r_{\pm}) (r_{3} - r_{\mp}) (r_{4} - r_{\pm})}
  {(r_{1} - r_{2}) (r_{2} - r_{3}) (r_{2} - r_{4}) (r_{+}-r_{-})}
  \,, \\
  \gamma_{r,0,\alpha} & \equiv
  - \frac{(r_{3} - r_{+})(r_{3} - r_{-}) (r_{1} + r_{2} - r_{3} + r_{4} - r_{+} - r_{-})}
  {(r_{1} - r_{3}) (r_{2} - r_{3}) (r_{2} - r_{4})}
  \,, \\
  \delta_{r,0,\alpha} &\equiv
  \frac{(r_{3}-r_{+})(r_{3}-r_{-})}{(r_{1}-r_{3})(r_{2}-r_{4})}
  \,.
\end{align}
Note the presence of some opposite signs $\mp$ in
Eq.~\eqref{eq:gamma_plusminus}.  After this decomposition, we only
need two types of integrals: one for the $\gamma$ terms, and one for
the $\delta$ term.  Namely, we have the definition of the incomplete
elliptic integral of the third kind (following the conventions
of the NIST Digital Library of Mathematical Functions~\cite{NIST:DLMF}),
\begin{align}
  \mathsf{\Pi}(\phi,\alpha^{2},k) = \int_{0}^{\sin\phi} \frac{d\rho}
  {(1-\alpha^{2}\rho^{2})\sqrt{(1-\rho^{2})(1-k^{2}\rho^{2})}}
  \,.
\end{align}
For $\phi=\pi/2$, $\sin\phi=1$, we have the complete elliptic integral
$\mathsf{\Pi}(\alpha^{2}, k) = \mathsf{\Pi}(\pi/2, \alpha^{2}, k)$.  Meanwhile, for the
$\delta$ term, we need the antiderivative given in Eq.~336.02 of Byrd
and Friedman~\cite{byrd2013handbook},
\begin{align}
  \label{eq:BF_336_02}
  \begin{split}
    & \int_{0}^{\sin\phi} \frac{d\rho}
    {(1-\alpha^{2}\rho^{2})^{2}\sqrt{(1-\rho^{2})(1-k^{2}\rho^{2})}} 
    \\
    &=
    \frac{1}{2(\alpha^{2}-1)(k^{2}-\alpha^{2})} \Big[ \alpha^{2}
    \mathsf{E}(\phi, k) + (k^{2}-\alpha^{2})u 
    \\
    &\; + (2\alpha^{2}k^{2}+2\alpha^{2}-\alpha^{4}-3k^{2})\mathsf{\Pi}(\phi,\alpha^{2},k)
    - \frac{\alpha^{4} \sn u \cn u \dn u}{1-\alpha^{2}\sn^{2} u}
    \Big] \!.
  \end{split}
\end{align}
Here we encounter the incomplete elliptic integrals of the first and second
kind, $u=\mathsf{F}(\phi,k)$ and $\mathsf{E}(\phi,k)$, defined by
\begin{align}
  \mathsf{F}(\phi,k) &= \int_{0}^{\sin\phi} \frac{d\rho}
  {\sqrt{(1-\rho^{2})(1-k^{2}\rho^{2})}}
  \,,\\
  \mathsf{E}(\phi,k) &= \int_{0}^{\sin\phi} \frac{\sqrt{1-k^{2}\rho^{2}}}
  {\sqrt{1-\rho^{2}}}d\rho
  \,.
\end{align}
We also encounter the Jacobi elliptic sine, cosine, and delta
amplitude functions $\sn, \cn, \dn$.  The Jacobi elliptic sine $\sn$
is kind of an inverse of $\mathsf{F}$ in the $\sin\phi$ argument, defined through the
relationship
\begin{align}
  u = \mathsf{F}(\phi, k) \quad \Longleftrightarrow \quad
  \sin\phi = \sn(u, k)
  \,.
\end{align}
It is customary to suppress the $k$ argument to $\sn,\cn,\dn$ and the
other Jacobi elliptic integrals when there is no possibility for
confusion.  The elliptic cosine and delta amplitude are related to the
elliptic sine through
\begin{align}
  &\sn^{2}u + \cn^{2} u = 1 \,, \\
  &(1-k^{2}) \sn^{2} u + \cn^{2} u =
  1 - k^{2}\sn^{2}u = \dn^{2} u
  \,.
\end{align}
The rich structure of elliptic functions is not necessary for our
purposes; we only need to evaluate the identity in
Eq.~\eqref{eq:BF_336_02} at $\phi=\pi/2$, where the elliptic integrals
become complete, with $\mathsf{K}(k) = F(\pi/2,k)$, $\mathsf{E}(k) = E(\pi/2,k)$; the
Jacobi sine evaluates $\sn u = \sin\phi=1$ at this point, so $\cn u =
0$.  This gives the identity we need,
\begin{align}
  \label{eq:BF_complete_double_pole_ident}
  \begin{split}
    & \int_{0}^{1} \frac{d\rho}
    {(1-\alpha^{2}\rho^{2})^{2}\sqrt{(1-\rho^{2})(1-k^{2}\rho^{2})}} 
    \\ & =
    \frac{1}{2(\alpha^{2}-1)(k^{2}-\alpha^{2})} \Big[ \alpha^{2}
    \mathsf{E}(k) + (k^{2}-\alpha^{2}) \mathsf{K}(k) 
    \\
    & \; + (2\alpha^{2}k^{2}+2\alpha^{2}-\alpha^{4}-3k^{2})\mathsf{\Pi}(\alpha^{2},k)
    \Big] \,.
  \end{split}
\end{align}
Adding together the three integrals multiplying
$\gamma_{+}, \gamma_{-}$, and $\gamma_{\alpha}$, and the above
identity for the $\delta_{\alpha}$ term, we finally arrive at eq. \eqref{eq:J_r_0_eval}.

When taking derivatives of this action integral with
respect to $E_{\rm so}, L_{\rm so}, K_{\rm so}$, it is
important to remember that each of $r_{1},r_{2},r_{3},r_{4}$, and the
combinations $k_{r}, \rho_{\pm}, \alpha_{r}$ depend on the constants
of motion; we discuss this issue further in Appendix \ref{app:freqs}.

\subsection{Radial spin correction}
We use the same transformation to $\rho$ as in Eq.~\eqref{eq:rho_def} to obtain
\begin{align}
\begin{split}
  & \frac{1}{\sqrt{R(r)}}
  \left[
    \sqrt{K_{\rm so}} \frac{ E_{\rm so} (r^2 + a^2) - a L_{\rm so}}{K_{\rm so}+ r^2}
  \right] dr
  \\
  & =
  \frac{2 \sqrt{K_{\rm so}}}{\sqrt{1-E_{\rm so}^{2}}\sqrt{(r_{1}-r_{3})(r_{2}-r_{4})}}
  \frac{\mathcal{R}_r(\rho) d\rho}{\sqrt{(1-\rho^{2})(1-k_{r}^{2}\rho^{2})}}
  \,, 
\end{split}
  \\
\begin{split}
  & \mathcal{R}_r(\rho^{2}) \equiv \frac{1}{\mathcal{D}} \Big(
        (a^2 E_{\rm so} - a L_{\rm so}) \left[(r_{1} - r_{2}) \rho^2-(r_{1} - r_{3})\right]^2 
        \\ 
        & \phantom{\mathcal{R}_r(\rho^{2}) \equiv}
        + E_{\rm so} \left[(r_{1} - r_{2}) r_{3} \rho^2-r_{2} (r_{1} - r_{3})\right]^2 \Big)
  \,,
\end{split} \\
\begin{split}
       & \mathcal{D} \equiv 
        K_{\rm so} \big[(r_{1} - r_{2}) \rho^2-(r_{1} - r_{3})\big]^2 + \big[(r_{1} - r_{2}) r_{3} \rho^2
        \\ 
        & \phantom{\mathcal{D} \equiv }
        -r_{2} (r_{1} - r_{3})\big]^2 \,.
\end{split}
\end{align}
Notice that both the numerator and denominator of the rational
polynomial $\mathcal{R}_r(\rho^{2})$ are quadratics in $\rho^{2}$, so they can be
factored into linear factors as
\begin{align}
  \mathcal{R}_r(\rho^{2}) =
  \frac{a^2 E_{\rm so} - a L_{\rm so} + E_{\rm so} r_{3}^2}{K_{\rm so} + r_{3}^2}
  \frac{(\rho^{2}-\rho_{z,+}^{2})(\rho^{2}-\rho_{z,-}^{2})}{(\rho^{2}-\rho_{p,+}^{2})(\rho^{2}-\rho_{p,-}^{2})}
  \,,
\end{align}
where we have defined the zeros and poles as
\begin{align}
\begin{split}
  \rho_{z,\pm}^{2} = &
  \frac{r_{1} - r_{3}}{r_{1} - r_{2}} \times
  \\
  & \frac{a^2 E_{\rm so} - a L_{\rm so} + E_{\rm so} r_{2} r_{3} \pm (r_{2} - r_{3})\sqrt{a E_{\rm so} (- a E_{\rm so} + L_{\rm so})} }
  {a^2 E_{\rm so} - a L_{\rm so} + E_{\rm so} r_{3}^2}
  \,,
\end{split}
  \\
  \rho_{p,\pm}^{2} &=
  \frac{r_{1}-r_{3}}{r_{1}-r_{2}}
  \frac{K_{\rm so} + r_{2} r_{3} \pm (r_{2}-r_{3}) \sqrt{- K_{\rm so}} }
  {K_{\rm so} + r_{3}^2}
  \,.
\end{align}
Now, performing a partial fractions decomposition, we find
\begin{align}
  \frac{(\rho^{2}-\rho_{z,+}^{2})(\rho^{2}-\rho_{z,-}^{2})}{(\rho^{2}-\rho_{p,+}^{2})(\rho^{2}-\rho_{p,-}^{2})}
  =
  \lambda
  + \frac{\gamma_{r,1,+}}{\rho^{2}-\rho_{p,+}^{2}}
  + \frac{\gamma_{r,1,-}}{\rho^{2}-\rho_{p,-}^{2}},
\end{align}
where the coefficients are $\lambda=1$ and
\begin{align}
  \gamma_{r,1,\pm} &= \pm \frac{(\rho_{p,\pm}^{2}-\rho_{z,+}^{2})(\rho_{p,\pm}^{2}-\rho_{z,-}^{2})}{\rho_{p,+}^{2}-\rho_{p,-}^{2}}
  \,.
\end{align}
So, the only types of integrals we need for $J_{r}^{(1)}$ are
\begin{align}
  \int_{0}^{1}
  \frac{d\rho}
  {\sqrt{(1-\rho^{2})(1-k_{r}^{2}\rho^{2})}}
  &=
  \mathsf{K}(k_{r})
  \,,\\
  \int_{0}^{1}
  \frac{\gamma_{r,1,\pm}}{\rho^{2}-\rho_{p,\pm}^{2}}
  \frac{d\rho}
  {\sqrt{(1-\rho^{2})(1-k_{r}^{2}\rho^{2})}}
  &=
  \frac{-\gamma_{r,1,\pm}}{\rho_{p,\pm}^{2}}
  \mathsf{\Pi}(\rho_{p,\pm}^{-2},k_{r})
  \,.
\end{align}
Putting it all together, we get the combination in eq.~\eqref{eq:J_r_1_eval},
where we have the two combinations 
\begin{align}
\begin{split}
  & \frac{\gamma_{r,1,\pm}}{\rho_{p,\pm}^{2}} 
  =
  \\ & \frac{(a^2 E_{\rm so} - a L_{\rm so} - E_{\rm so} K_{\rm so}) (r_{2} - r_{3}) (K_{\rm so} - r_{3}(r_{3} \pm 2 \sqrt{-K_{\rm so}}))}
  {2 (a^2 E_{\rm so} - a L_{\rm so} + E_{\rm so} r_{3}^2)(K_{\rm so}(r_{2}-r_{3}) \mp (K_{\rm so}+r_{2}r_{3})\sqrt{-K_{\rm so}})}
  .
\end{split}
\end{align}
Despite the fact that these combinations are complex (since
$\sqrt{-K_{\rm so}}$ is imaginary), the $\pm$ pairs are complex conjugates,
as are $\rho_{p,\pm}^{2}$.  As can be seen from the definition, when
$k_{r}^{2}$ is real and $k_{r}^{2}<1$ (as is true for bound stable orbits), we have
\begin{align}
  \mathsf{\Pi}\left(\overline{\alpha^{2}},k\right)
  =
  \overline{\mathsf{\Pi}(\alpha^{2},k)}
  \,,
\end{align}
where the overline denotes complex conjugation.  Since in
Eq.~\eqref{eq:J_r_1_eval} we then have the sum of two complex conjugate
pairs, the action is real.

\subsection{Polar geodesic action}
Recall that the polar potential can be factored as
\begin{align}
\begin{split}
  Z_{\rm so}(z) =
    &  [K_{\rm so}-(a E_{\rm so}-L_{\rm so})^{2}] (1-z^{2}) 
    \\
    & - z^{2} [a^{2}(1-E_{\rm so}^{2})(1-z^{2})+L_{\rm so}^{2}]
    \\
    =&  a^{2}(1-E_{\rm so}^{2})(z_{2}^{2}-z^{2})(z_{1}^{2}-z^{2})
   \,.
\end{split}
\end{align}
Here $\pm z_{2}$ are the two physical
turning points of the polar motion, and $\pm z_{1}$ are two other
roots of the quartic, which can be found in terms of $a,E_{\rm so},z_{2}$
by depressing the quartic by $(z_{2}^{2}-z^{2})$, yielding
\begin{align}
  z_{1}^{2} =
  1-z_{2}^{2}+\frac{L_{\rm so}^2 + K_{\rm so} - (a E_{\rm so} - L_{\rm so})^2}{a^2 (1 - E_{\rm so}^2)}
  \,.
\end{align}
Notice, by matching coefficients, the simple expression for
$z_{2}^{2}+z_{1}^{2}$ (above), and also for the product
$z_{2}^{2}z_{1}^{2}$,
\begin{align}
  z_{2}^{2}z_{1}^{2} = \frac{K_{\rm so} - (a E_{\rm so} - L_{\rm so})^2}{a^{2}(1-E_{\rm so}^{2})}
  \,.
\end{align}

Now we return to our goal integral,
\begin{align}
  J_{z}^{(0)} = \frac{2}{\pi} \int_{0}^{z_{2}} \frac{\sqrt{Z(z)}}{1-z^{2}} dz
  \,.
\end{align}
As usual, we want to put the integrand in the Legendre standard form,
with the radical of a quartic in the denominator, and remap the poles
at $\pm z_{2}$ to $\pm 1$ in a new coordinate system.  We only need
the linear transformation
\begin{align}\label{eq:zeta_def}
  \zeta &= \frac{z}{z_{2}} \,,
\end{align}
and our integral becomes
\begin{align}
\begin{split}
  J_{z}^{(0)} = & \frac{2a\sqrt{1-E_{\rm so}^{2}}z_{2}^{2}z_{1}}{\pi} \times
  \\
  & \int_{0}^{1} \frac{(1-\zeta^{2})(1-k_{z}^{2}\zeta^{2})}{(1-z_{2}^{2}\zeta^{2})\sqrt{(1-\zeta^{2})(1-k_{z}^{2}\zeta^{2})}} d\zeta\,.
\end{split}
\end{align}
where we have defined
\begin{align}
  k_{z}^{2} \equiv \frac{z_{2}^{2}}{z_{1}^{2}}
  \,.
\end{align}
Now we use the partial fractions decomposition in factors of
$\zeta^{2}$,
\begin{align}
  \frac{(1-\zeta^{2})(1-k_{z}^{2}\zeta^{2})}{1-z_{2}^{2}\zeta^{2}}
  =
  \beta_{z,0} + \eta \zeta^{2} + \frac{\gamma_{z,0}}{1-z_{2}^{2}\zeta^{2}}
  \,,
\end{align}
where the coefficients are
\begin{align}
  \beta_{z,0} &= \frac{-1 + z_{2}^2 + z_{1}^2}{z_{2}^2 z_{1}^2}
  \,, &
  \eta &= \frac{-1}{z_{2}^{2}}
  \,, &
  \gamma_{z,0} &= \frac{(1-z_{2}^{2})(1-z_{1}^{2})}{z_{2}^{2}z_{1}^{2}}
  \,.
\end{align}
We have already seen the elliptic integrals $\mathsf{K}$ and $\mathsf{\Pi}$
defined above, which take care of the $\beta_{0}$ and $\gamma_{z,0}$
terms; the only other one we need is decomposed as a combination of
the first and second kind,
\begin{align}
  \int_{0}^{1} \frac{\zeta^{2}\ d\zeta}
  {\sqrt{(1-\zeta^{2})(1-k_{z}^{2}\zeta^{2})}}
  =
  \frac{\mathsf{K}(k_{z})-\mathsf{E}(k_{z})}{k_{z}^{2}}
  \,.
\end{align}
Adding together the three terms, we get the simple expression in Eq.\eqref{eq:J_z_0_eval}.

\subsection{Polar spin correction}

Transforming to the same coordinate as at geodesic order $\zeta=z/z_{-}$, the $J_z^{(1)}$ integrals from Eq. \eqref{eq:J_z_1_def} becomes
\begin{align}
\begin{split}
  J_{z}^{(1)} =
      & \frac{-2s_{\parallel}}{\pi} \int_{0}^{1}
      \frac{\sqrt{K_{\rm so}}}{z_{+}\sqrt{1-E_{\rm so}^{2}}}
      \frac{1}{\sqrt{(1-\zeta^{2})(1-k_{\theta}^{2}\zeta^{2})}} \times
      \\ 
      &\frac{L_{\rm so}-a E_{\rm so}(1-z_{-}^{2}\zeta^{2})}{K_{\rm so}-a^{2}z_{-}^{2}\zeta^{2}} d\zeta
  \,.
\end{split}
\end{align}
Once more we perform a partial fractions decomposition in terms of
$\zeta^{2}$,
\begin{align}
  \frac{L_{\rm so}-a E_{\rm so}(1-z_{-}^{2}\zeta^{2})}{K_{\rm so}-a^{2}z_{-}^{2}\zeta^{2}}
  =
  \beta_{1} + \frac{\gamma_{z,1}}{1-\alpha_{z}^{2}\zeta^{2}}
  \,,
\end{align}
where we define
\begin{align}
  \beta_{1} &= -\frac{E_{\rm so}}{a}
  \,, &
  \alpha_{z}^{2} &= \frac{a^{2}z_{-}^{2}}{K_{\rm so}}
  \,, &
  \gamma_{z,1} &=\frac{E_{\rm so}(K_{\rm so}-a^{2}) + a L_{\rm so}}{a K_{\rm so}}
  \,.
\end{align}
So we immediately integrate the definitions of $K$ and $\Pi$ and obtain eq. \eqref{eq:J_z_1_eval}.

\section{Derivatives of actions and frequencies} \label{app:freqs}

In this appendix, we describe the calculation of the frequencies from the derivatives of the actions. Similarly to the previous section of the appendix, we drop the ``so'' and ``c'' subscripts on the constants of motion since $C_c = C_\text{so} + \mathcal{O}(s,s_\parallel)$.

The integrals for the derivatives of $J_r^{(1)}$ in Eqs.~\eqref{eq:dJr1dE} to \eqref{eq:dJr1dK} and of $J_z^{(1)}$ in Eqs.~\eqref{eq:dJz1dE} to \eqref{eq:dJz1dK} require Hadamard's partie finie regularization. It can be calculated as follows.

\subsection{Derivatives of the correction to the radial action}

First, the radial integrals can be decomposed into partial fractions in $r$ as
\begin{multline}
    \int_{r_2}^{r_1} \Bigg( C_0 + \sum_{i=1}^4 \frac{C_i}{r-r_i} + \frac{C_{-i\sqrt{K}}}{r-i\sqrt{K}} + \frac{D_{-i\sqrt{K}}}{(r-i\sqrt{K})^2} \\ + \frac{C_{i\sqrt{K}}}{r+i\sqrt{K}} + \frac{D_{i\sqrt{K}}}{(r+i\sqrt{K})^2} \Bigg)\frac{\d r}{\sqrt{R(r)}} \,,
\end{multline}
where $C_i$, $C_{\pm i\sqrt{K}}$ and $D_{\pm i\sqrt{K}}$ are constants specific for each derivative. The integrals with $C_0$, $C_3$, $C_4$, $C_{\pm i\sqrt{K}}$, and $D_{\pm i\sqrt{K}}$ are regular and no partie finie is needed. After transformation to $\rho$ according to Eq.~\eqref{eq:rho_def}, they can be calculated using the standard techniques presented in Appendix \ref{app:leg}.

However, the integrals with $C_{1,2}$ diverge with the power of $-3/2$ in one of the integration bounds and must be calculated in the sense of Hadamard's partie finie. After the substitution \eqref{eq:rho_def}, the integrals can be written as
\begin{widetext}
\begin{align}
    \int_{r_2}^{r_1} \frac{\d r}{(r-r_1)\sqrt{R(r)}} &= - \frac{\mathcal{C}_r}{r_1-r_3} \int_0^1 \left( 1 + \frac{r_2-r_3}{(r_1-r_2)(1-\rho^2)} \right) \frac{\d \rho}{\sqrt{(1-\rho^2)(1-k_r^2 \rho^2)}} \,, \\
    \int_{r_2}^{r_1} \frac{\d r}{(r-r_2)\sqrt{R(r)}} &= - \frac{\mathcal{C}_r}{r_2-r_3} \int_0^1 \left( 1 - \frac{r_1-r_3}{(r_1-r_2)\rho^2} \right) \frac{\d \rho}{\sqrt{(1-\rho^2)(1-k_r^2 \rho^2)}}\,,
\end{align}
\end{widetext}
where
\begin{equation}
    \mathcal{C}_r = \frac{2}{\sqrt{(1-E^2)(r_1-r_3)(r_2-r_4)}} \,.
\end{equation}
The first term in each bracket can be expressed as $\mathsf{K}(k_{r})$. The second term has a singular and regular part and must be regularized. This can be done by changing the integration bounds to $(0,1-\epsilon)$ in the first integral and to $(\epsilon,1)$ in the second integral. Then, they can be expressed using incomplete elliptic integrals and expanded in Laurent series around $\epsilon = 0$. The results then read
\begin{multline}\label{eq:integral_div_1}
    \int_0^{1-\epsilon} \frac{\d \rho}{(1-\rho^2)^{3/2} \sqrt{1-k^2 \rho^2}} = \frac{1}{\sqrt{2(1-k^2)\epsilon}} \\ + \mathsf{K}(k) - \frac{\mathsf{E}(k)}{1-k^2} + \mathcal{O}(\sqrt{\epsilon}) \,,
\end{multline}
\begin{equation}
    \int_\epsilon^1 \frac{\d \rho}{\rho^2 \sqrt{(1-\rho^2)(1-k^2 \rho^2)}} = \frac{1}{\epsilon} + \mathsf{K}(k) - \mathsf{E}(k) + \mathcal{O}(\epsilon) \,.
\end{equation}
We can see that both integrals contain a divergent part proportional to $\epsilon^{-1/2}$ and $\epsilon^{-1}$, respectively. The Hadamard partie finie is the second term in the series which is regular. Putting everything together, we obtain
\begin{equation}
    \int_{r_2}^{r_1} \frac{\d r}{(r-r_1)\sqrt{R(r)}} = \frac{\mathcal{C} ((r_2-r_4)\mathsf{E}(k_r) - (r_1-r_4)\mathsf{K}(k_r))}{(r_1-r_2)(r_1-r_4)} \,,
\end{equation}
\begin{multline}
    \int_{r_2}^{r_1} \frac{\d r}{(r-r_2)\sqrt{R(r)}} = \frac{\mathcal{C}((r_2-r_3)\mathsf{K}(k_r) - (r_1-r_3)\mathsf{E}(k_r))}{(r_1-r_2)(r_2-r_3)} \,.
\end{multline}

The final formulas for $\partial_{E,L,K} J_r^{(1)}$ can be expressed as
\begin{widetext}
\begin{subequations}\label{eq:Jr1_derivatives_results}
\begin{align}
    \frac{\partial J_r^{(1)}}{\partial E} &= \frac{s_\parallel \mathcal{C}_r \sqrt{K}}{\pi(1-E^2)} \left( \mathsf{K}(k_r) + \sum_{i=1}^4 (a^2+r_i^2) \Delta(r_i) I_i^r \prod_{\substack{j=1 \\ j \neq i}}^4 \frac{1}{r_i - r_j} \right) \,, 
    \\
    \frac{\partial J_r^{(1)}}{\partial L} &= -\frac{s_\parallel a \mathcal{C}_r \sqrt{K}}{\pi(1-E^2)} \sum_{i=1}^4 \Delta(r_i) I_i^r \prod_{\substack{j=1 \\ j \neq i}}^4 \frac{1}{r_i - r_j} \,,
    \\
    \frac{\partial J_r^{(1)}}{\partial K} &= -\frac{s_\parallel \mathcal{C}_r}{2 \sqrt{K} \pi} \Bigg( - E \mathsf{K}(k_r) + \frac{K}{1-E^2} \sum_{i=1}^4 \frac{((a^2+r_i^2) E - a L)\Delta(r_i)}{K + r_i^2} I_i^r \prod_{\substack{j=1 \\ j \neq i}}^4 \frac{1}{r_i - r_j} + \frac{(1-E^2)(r_1-r_3)(r_2-r_4)}{2((a^2-K)E - a L)} \mathsf{E}(k_r) \nonumber\\
    &\phantom{=} + \left( \frac{K( (a^2 - K)(1 - 2 E^2) + 2 a E L - 2 M r_3  )}{(K + r_3^2)((K-a^2) E + a L)}  + \frac{((a^2-K)E-aL)(K - r_3^2)}{(K+r_3^2)^2} \right) \mathsf{K}(k_r) + \nonumber \\
    &\phantom{=} + \frac{(1-E^2)^2 (r_1-r_3) (r_2-r_3) ( K^2 - K( r_1 r_2 + 2(r_1+r_2)r_3 + r_3^2 ) + r_1 r_2 r_3^2 )(K+r_4^2)}{2((a^2-K)E - a L)^3 (K+r_3^2)} \mathsf{K}(k_r) \Bigg)\,,
\end{align}
\end{subequations}
\end{widetext}
where
\begin{align}
    I_1^r &= \frac{(r_2-r_4) \mathsf{E}(k_r) - (r_1-r_4) \mathsf{K}(k_r)}{(r_1-r_2)(r_1-r_4)} \,, 
    \\
    I_2^r &= \frac{-(r_1-r_3) \mathsf{E}(k_r) + (r_2-r_3) \mathsf{K}(k_r)}{(r_1-r_2)(r_2-r_3)} \,,
    \\
    I_3^r &= \frac{(r_2-r_4) \mathsf{E}(k_r) - (r_2-r_3) \mathsf{K}(k_r)}{(r_2-r_3)(r_3-r_4)} \,,
    \\
    I_4^r &= \frac{-(r_1-r_3) \mathsf{E}(k_r) + (r_1-r_4) \mathsf{K}(k_r)}{(r_3-r_4)(r_1-r_4)} \,.
\end{align}
Interestingly, all the terms containing elliptic integrals with complex arguments subtract, which makes the final expressions manifestly real.

\subsection{Derivatives of the correction to the polar action}

Similarly to the radial action, we can decompose the integrals \eqref{eq:dJz1dE} -- \eqref{eq:dJz1dK} into partial fractions with respect to $z^2$ as
\begin{equation}
\begin{split}
    & \int_0^{z_2} \frac{\d z}{\sqrt{Z(z)}} \bigg( C_0 + \frac{C_1}{z^2 - z_1^2} + \frac{C_2}{z^2 - z_2^2} + \frac{C_K}{K - a^2 z^2} 
    \\ 
    & \phantom{\int_0^{z_2} \frac{\d z}{\sqrt{Z(z)}} \bigg(}
    + \frac{D_K}{(K - a^2 z^2)^2} \bigg) \,.
\end{split}
\end{equation}
After using the substitution \eqref{eq:zeta_def}, the terms with $C_0$, $C_1$, $C_K$ and $D_K$ are regular and can be calculated using standard techniques. The term with $C_2$ is of the type \eqref{eq:integral_div_1} and diverges in $1$. Again, by using its partie finie, we can express the derivatives of the polar action, which read
\begin{widetext}
\begin{subequations}\label{eq:Jz1_derivatives_results}
\begin{align}
    \frac{\partial J_z^{(1)}}{\partial E} &= \frac{2 s_\parallel \sqrt{K} \left((z_{2}^2-z_{1}^2) \left(z_{2}^2 \left(z_{1}^2-2\right)+1\right) \mathsf{K}(k_z)+\left(z_{2}^2 \left(\left(z_{2}^2-4\right) z_{1}^2+z_{1}^4+1\right)+z_{1}^2\right) \mathsf{E}(k_z)\right)}{\pi  a \left(1-E^2\right)^{3/2} z_{1} z_2^2 \left(z_{2}^2-z_{1}^2\right)^2} \,,
    \\
    \frac{\partial J_z^{(1)}}{\partial L} &= \frac{2 s_\parallel \sqrt{K} \left(\left(z_{2}^2-1\right) (z_{2}^2-z_{1}^2) \mathsf{K}(k_z)+\left(z_{2}^2 \left(2 z_{1}^2-1\right)-z_{1}^2\right) \mathsf{E}(k_z)\right)}{\pi  a^2 \left(1-E^2\right)^{3/2} z_{2}^2 z_{1} \left(z_{2}^2-z_{1}^2\right)^2} \,,
    \\
    \frac{\partial J_z^{(1)}}{\partial K} &= -\frac{s_\parallel}{\pi a z_1 \sqrt{K (1-E^2)}} \bigg( E \mathsf{K}(k_z) - \frac{K}{a (1-E^2) (z_2^2-z_1^2)} \sum_{i=1}^2 \frac{(1-z_i^2)(L - a E (1-z_i^2))}{z_i^2 (K - a^2 z_i^2)} I^z_i \nonumber \\ 
    &\phantom{=} + \frac{((a^2-K)E - a L) ( a^2 z_1^2 \mathsf{E}(k_z) + (K - a^2 z_1^2) \mathsf{K}(k_z) )}{(K - a^2 z_1^2)(K - a^2 z_2^2)} \bigg) \,.
\end{align}
\end{subequations}
\end{widetext}
where
\begin{align}
    I_1^z &= - \frac{\mathsf{E}(k_z)}{1-k_z^2} \,, 
    \\
    I_2^z &= \mathsf{K}(k_z) - \frac{\mathsf{E}(k_z)}{1-k_z^2}\,.
\end{align}
Again, the terms that contain elliptic integrals of the third kind $\mathsf{\Pi}(a^2 z_2^2/K, k_z)$ subtract, simplifying the final expression.

\subsection{Numerical verification of frequencies}

\begin{figure}
    \centering
    \includegraphics[width=\linewidth]{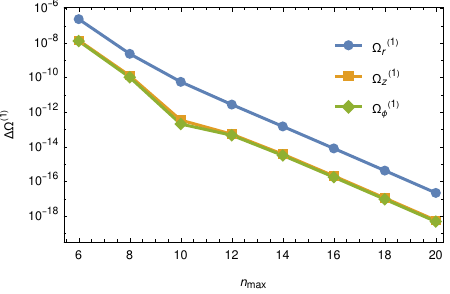}
    \caption{Difference between the frequencies calculated using the analytical formulas derived in this work and the numerical calculation derived in \cite{Drummond:2022efc,Drummond:2022xej}. $n_\text{max}$ on the horizontal axis denotes number of Fourier coefficients in the numerical calculation. As this number increases, the difference converges to zero. The parameters of the orbit are $a=0.9M$, $E_\text{so} = 0.9543$, $L_\text{so} = 3.1386M$ and $K_\text{so} = 4.6041M^2$.}
    \label{fig:comparisons_DH}
\end{figure}

To verify the analytical formulas for the frequencies, we compared them with the frequencies calculated using the approach described in \cite{Drummond:2022efc,Drummond:2022xej}. In this approach, the equations of motion are expanded in a Fourier series and solved for the Fourier coefficients as a system of linear equations. Therefore, there exists a numerical error caused by the truncation of the Fourier series. Thus, we calculate the frequencies for several numbers of Fourier coefficients to verify the convergence. In this way, we compared the frequencies at several points of the parameter space and in Figure \ref{fig:comparisons_DH} we show the plot of the differences between the linear parts of the coordinate frequencies $\Omega_r^{(1)}$, $\Omega_z^{(1)}$, $\Omega_\phi^{(1)}$ calculated analytically and using the Fourier expansion\footnote{The linear parts of the frequencies calculated by \citet{Drummond:2022efc,Drummond:2022xej} are calculated with respect to a geodesic with fixed average turning points as opposed to this work where we fix the constants of motion. Therefore, we have to additionally transform between these parametrizations.} for an orbit with $a=0.9M$, $E_\text{so} = 0.9543$, $L_\text{so} = 3.1386M$ and $K_\text{so} = 4.6041M^2$. We can see that with an increasing number of Fourier coefficients $n_\text{max}$ the difference converges to zero.

\subsection{The deformed Carter-Mino time}

The frequencies with respect to Carter-Mino time $\mathrm{d} \lambda = \Sigma^{-1} \mathrm{d}\tau$ can be calculated in the geodesic case by using the Mino-time Hamiltonian $H_\lambda = \Sigma (H_\tau + 1/2 )$. In the expressions for the frequencies \eqref{eq:frequencies_Schmidt} the derivatives with respect to the Hamiltonian can be calculated as 
\begin{equation}
    \frac{\partial J_r}{\partial H_\lambda} = - \frac{\partial J_r}{\partial K} \, , \quad \frac{\partial J_z}{\partial H_\lambda} = \frac{\partial J_z}{\partial K} ,
\end{equation}
which leads to the Mino frequencies
\begin{align}
    \Upsilon^r &= - \frac{1}{2} \left( \frac{\partial J_r}{\partial K} \right)^{-1} , \\
    \Upsilon^z &=  \frac{1}{2} \left( \frac{\partial J_z}{\partial K} \right)^{-1} , \\
    \Upsilon^t &= \frac{1}{2} \left( - \frac{\partial J_r}{\partial E} \left( \frac{\partial J_r}{\partial K} \right)^{-1} + \frac{\partial J_z}{\partial E} \left( \frac{\partial J_z}{\partial K} \right)^{-1} \right) , \\
    \Upsilon^\phi &= \frac{1}{2} \left( \frac{\partial J_r}{\partial L} \left( \frac{\partial J_r}{\partial K} \right)^{-1} - \frac{\partial J_z}{\partial L} \left( \frac{\partial J_z}{\partial K} \right)^{-1} \right) , \\
    \Upsilon^\psi &= \frac{1}{2} \left( \frac{\partial J_r}{\partial s_\parallel} \left( \frac{\partial J_r}{\partial K} \right)^{-1} - \frac{\partial J_z}{\partial s_\parallel} \left( \frac{\partial J_z}{\partial K} \right)^{-1} \right) .
\end{align}
We can repeat this process for the linear-in-spin parts of the actions to obtain frequencies of the trajectory of spinning bodies. However, these frequencies do not match the Carter-Mino time frequencies calculated in \cite{Drummond:2022xej,Piovano:2024yks}. This is due to the fact that the frequencies above are not with respect to Carter-Mino time in the spinning case. Instead, they are defined with respect to a parameter $\Tilde{\lambda}$ which differs from Carter-Mino time as
\begin{equation}
    \left\langle \frac{\d \lambda}{\d \Tilde{\lambda}} \right\rangle = 1 + s_\parallel \delta \dot{\lambda}\,,
\end{equation}
where the angle brackets denote averaging over the radial and polar motion and where
\begin{equation}
    \delta \dot{\lambda} = \left\langle \frac{((r^2 - a^2 z^2) E - 2a(L - a E) )(r^2 - a^2 z^2)}{\sqrt{K} (r^2 + a^2 z^2)^2} \right\rangle\,.
\end{equation}
This formula was found empirically by comparing the numerically evaluated frequencies calculated with different methods. The expression for $\delta \lambda$ is not separable to a sum of functions of only $r$ and only $z$, making it impossible to express using Legendre elliptic integrals using the standard approach for calculating geodesic averages.

\section{Details of PN expansion of actions}
\label{app:PNexp}

We used the expansion of elliptic integrals \cite{WolframElliptic}
\begin{align}
  & \mathsf{K}(k) = \frac{\pi}{2} \sum_{i=0}^\infty \frac{[(1/2)_i]^2  k^{2i}}{(i!)^2}\,,
  \\
  & \mathsf{E}(k) = \frac{\pi}{2} \sum_{i = 0}^\infty \frac{(-1/2)_i (1/2)_i k^{2i}}{(i!)^2}\,,
  \\
\begin{split}
  & \mathsf{\Pi}(n,k) = \frac{\pi}{2} \sum_{i=0}^\infty \frac{[(1/2)_i]^2 }{(i!)^2} 
  \Bigg[ \frac{i!}{n^i \sqrt{1-n} (1/2)_i} 
  \\
  & \phantom{\mathsf{\Pi}(n,k) =} - \frac{2i}{n} \sum_{j=0}^{i-1}\frac{(1-i)_j}{(3/2)_j}\left(1 - \frac{1}{n}\right)^j \Bigg] k^{2i} \,,
\end{split}
\end{align}
where $(a)_i$ is the Pochhammer symbol. These series representations were useful because the argument $k$ always turns out to be small in the PN limits of formulas \eqref{eq:J_r_0_eval}--\eqref{eq:J_z_1_eval}.

\subsection{Radial part}
%
Using the variables introduced in \eqref{eq:ellYe} one can easily expand the roots $r_{1,2,3,4}$ by iterating $R(r) = 0$ as
\begin{align}
\begin{split}
 & r_1 = 
  \frac{1}{\epsilon^2} \frac{(1 + \sqrt{1-\tilde{e}^2})\ell^2}{M \tilde{e}} 
  - \frac{1}{\epsilon} \frac{a (1 + \sqrt{1-\tilde{e}^2})Y \ell}{M(1 - \tilde{e}^2 + \sqrt{1 - \tilde{e}^2})} 
  \\ 
  & \phantom{r_1 =} 
  + \frac{M[(10 -\tilde{e}^2) \sqrt{1-\tilde{e}^2} - 6(1-\tilde{e}^2)]}{8(1-\tilde{e}^2)} 
  \\ 
  & \phantom{r_1 =}
  + \frac{a^2(1 - \tilde{e}^2(1 + Y^2))}{2(1-\tilde{e}^2)^{3/2}}
  + \mathcal{O}(\epsilon) \,,
\end{split}
 \\
\begin{split}
 & r_2 = 
  \frac{1}{\epsilon^2}  \frac{(1 - \sqrt{1-\tilde{e}^2})\ell^2}{M \tilde{e}}
  - \frac{1}{\epsilon} \frac{a (1 - \sqrt{1-\tilde{e}^2})Y \ell}{M(1 - \tilde{e}^2 - \sqrt{1 - \tilde{e}^2})} 
  \\ 
  & \phantom{r_1 =}
  - \frac{M[(10 -\tilde{e}^2) \sqrt{1-\tilde{e}^2} + 6(1-\tilde{e}^2)]}{8(1-\tilde{e}^2)} 
  \\ 
  & \phantom{r_1 =}
  - \frac{a^2(1 - \tilde{e}^2(1 + Y^2))}{2(1-\tilde{e}^2)^{3/2}} + \mathcal{O}(\epsilon)\,,
\end{split}
 \\
 & r_3 = M+\sqrt{M^2 - a^2(1-Y^2)} + \mathcal{O}(\epsilon)\,,
 \\
 & r_4 = M-\sqrt{M^2 - a^2(1-Y^2)} + \mathcal{O}(\epsilon) \,.
\end{align}
Note that $r_{3,4}$ start only at $\mathcal{O}(1)$, 1PN beyond leading order, which means that a naive expansion of $R(r)=0$ and an iteration of the leading order roots would miss these solution branches. Much of the difficulties with expansions of these formulas to high order came from simplifications of various polynomials of $\sqrt{1 - \tilde{e}^2}$ in the expressions. Indeed, the simplification of the formulas related to the radial roots and of the various expressions in which they appeared was the main bottleneck of the computations.
Formulas for the roots up to 3PN relative order are in the Supplementary notebook. By substituting these root expansions along with the transformation \eqref{eq:ellYe} into formulas \eqref{eq:J_r_0_eval}, \eqref{eq:J_r_1_eval} for $J_r^{(0,1)}$ we obtained the radial action expansions in equations \eqref{eq:J_r_0_PNexp} and \eqref{eq:J_r_1_PNexp}.

\subsection{Polar part}
The roots of $Z(z) = 0$ have a simple form non-perturbatively since $Z$ can be viewed as a quadratic polynomial in $z^2$. They can be directly expanded as
\begin{align}
\begin{split}
  & z^2_{1} = (1-Y^2) + \epsilon \frac{2 a Y^3}{\ell} 
  - \epsilon^2\frac{a^2 Y^2 (1+4 Y^2)}{\ell^2}
  \\
  & \phantom{z^2_{1} =}
  + \epsilon^3 \frac{a Y^3 [(a^2(4+8 Y^2)) - M^2 \tilde{e}^2]}{\ell^3} + \mathcal{O}(\epsilon^4) \,,
\end{split}
  \\
\begin{split}
  & z^2_2 = 
    \frac{1}{\epsilon^4}\frac{\ell^4}{M^2 a^2 \tilde{e}^2 } 
    + \frac{1}{\epsilon^3} \frac{2 Y \ell^3}{M^2 a \tilde{e}^2}
    + \frac{1}{\epsilon^2} \left(\frac{1}{4 a^2} - \frac{1}{M^2 \tilde{e}^2}\right)\ell^2
    \\
    & \phantom{z^2_2 =}
    - \frac{1}{\epsilon} \frac{Y \ell}{2 a}
    + \mathcal{O}(\epsilon^0)\,.
\end{split}
\end{align}
Again, substituting these expressions into the exact actions \eqref{eq:J_z_0_eval} and \eqref{eq:J_z_1_eval} led us to the expanded expressions \eqref{eq:J_z_0_PNexp} and \eqref{eq:J_z_1_PNexp}.

To invert the expressions, we resubstituted $\tilde{e} = \ell \sqrt{E_{\rm so} - 1} $ and $Y = L_{\rm so}/\ell = J_\phi / \ell$ and then eliminated $\ell = \sqrt{K_{\rm so}}$ from the expressions. By inverting the $J_z = J_z^{(0)}+J_z^{(1)} $ relation we obtained
\begin{align}
  \begin{split}
  & \ell = 
    J_z + |J_\phi| 
    - \epsilon \frac{a J_\phi}{J_z + |J_\phi|}
    + \epsilon^2 \frac{a^2 \left[(J_z +|J_\phi|)^2 - J_\phi^2\right]}{2 (J_z +|J_\phi|)^3}
    \\
    & \phantom{\ell =}
    + \epsilon^2 \frac{a s_\parallel J_\phi}{(J_z +|J_\phi|)^2}
    - \epsilon^3 \frac{a (E_{\rm so}-1) J_\phi}{J_z +|J_\phi|}
    \\
    & \phantom{\ell =}
    + \epsilon^3 \frac{a^3 J_\phi \left((J_z +|J_\phi|)^2 - J_\phi^2\right) }{2(J_z +|J_\phi|)^5}
    \\
    & \phantom{\ell =}
    +\epsilon ^3 \frac{s_\parallel a^2\left[(J_z +|J_\phi|)^2 - 3 J_\phi^2\right]}{2(J_z +|J_\phi|)^4} + \mathcal{O}(\epsilon^4)\,.
  \end{split}
\end{align}
We then plugged these expressions for $\ell$ into $J_r = J_r^{(0)} + J_r^{(1)}$ and inverted for $H_{\rm Ks} = E_{\rm so}-1$ perturbatively to obtain the Hamiltonian in equation \eqref{eq:HKs}.

\renewcommand{\leftmark}{List of references}
\bibliography{apssamp}

\end{document}